\documentclass[aps,prd,twocolumn,showpacs,superscriptaddress,nofootbib,floatfix,longbibliography, preprintnumbers]{revtex4-1}

\usepackage[utf8]{inputenc}
\usepackage{amsmath,amssymb,amsfonts}
\usepackage{mathrsfs}
\usepackage{graphicx}
\usepackage[dvipsnames]{xcolor}
\usepackage{bm}
\usepackage{physics}
\usepackage{hyperref}
\usepackage{orcidlink}
\usepackage[capitalise]{cleveref}
\usepackage{enumitem}
\usepackage{booktabs}
\usepackage{placeins}

\hypersetup{
    colorlinks=true,
    linkcolor=NavyBlue,
    citecolor=NavyBlue,
    urlcolor=NavyBlue
}

\setlist[description]{leftmargin=\parindent,labelindent=\parindent}

\newcommand{\mrm}[1]{\mathrm{#1}}

\newcommand{\nl}[1]{\nonumber \\}
\newcommand{\Pcas}{P_{\rm C}}
\newcommand{\Ppatch}{P_{\rm patch}}
\newcommand{\Pnew}{P_{\rm ps}}
\newcommand{\Ptot}{P_{\rm meas}}

\begin{document}

\preprint{KEK-QUP-2026-0020, KEK-TH-2887}

\title{Casimir Experiments as Spin-Force Detectors} 

\author{Xiaolin Ma~\orcidlink{0009-0007-1994-9493}}
\altaffiliation{\href{mailto:xlmphy@post.kek.jp}{xlmphy@post.kek.jp, corresponding}}
\affiliation{International Center for Quantum-field Measurement Systems for Studies of the Universe and Particles (QUP, WPI),
High Energy Accelerator Research Organization (KEK), Oho 1-1, Tsukuba, Ibaraki 305-0801, Japan}

\author{Volodymyr Takhistov}
\altaffiliation{\href{mailto:vtakhist@post.kek.jp}{vtakhist@post.kek.jp, corresponding}}
\affiliation{International Center for Quantum-field Measurement Systems for Studies of the Universe and Particles (QUP, WPI),
High Energy Accelerator Research Organization (KEK), Oho 1-1, Tsukuba, Ibaraki 305-0801, Japan}
\affiliation{Theory Center, Institute of Particle and Nuclear Studies (IPNS), High Energy Accelerator Research Organization (KEK), Tsukuba 305-0801, Japan
}
\affiliation{Graduate University for Advanced Studies (SOKENDAI),
1-1 Oho, Tsukuba, Ibaraki 305-0801, Japan}
\affiliation{Kavli Institute for the Physics and Mathematics of the Universe (WPI), UTIAS, \\The University of Tokyo, Kashiwa, Chiba 277-8583, Japan}

\author{Hideo Iizuka} 
\affiliation{International Center for Quantum-field Measurement Systems for Studies of the Universe and Particles (QUP, WPI),
High Energy Accelerator Research Organization (KEK), Oho 1-1, Tsukuba, Ibaraki 305-0801, Japan}

\begin{abstract}
Precision Casimir experiments enable highly sensitive probes of
short range forces, but have thus far constrained only
spin-independent interactions. We show that measurements between reversed polarizations of a magnetically compensated
source enable direct searches of axion-mediated
monopole-dipole forces. The spin force reverses sign, while the common Casimir, gravitational  and 
electrostatic backgrounds broadly cancel. Compensated sources need no intervening shielding, allowing large
probes at micrometer gaps. Casimir experiments
could probe the couplings $g_p^e g_s^{N}$ and $g_p^e g_s^{e}$ at
$\lambda\sim1-100~\mu{\rm m}$ orders of magnitude beyond existing 
laboratory searches.
\end{abstract}

\maketitle
{\it Introduction.-}
Searches for new macroscopic forces provide direct low energy probes of physics
beyond the Standard Model (SM). Such forces can arise over length scales associated with reduced Compton wavelength $\lambda = 1/m_\phi$ of a mediator with mass $m_\phi$, and can be generated by light, weakly coupled particles that are challenging to probe at high energy colliders. Particularly well motivated are axions and axion-like particles (collectively, axions), arising in connection with   the strong-CP problem of quantum chromodynamics, as dark matter candidates and generically occur in theories extending SM~\cite{
Peccei:1977hh,Peccei:1977ur,Weinberg:1977ma,
Wilczek:1977pj,Shifman:1979if,Zhitnitsky:1980tq,Dine:1981rt,Witten:1984dg,Arvanitaki:2009fg,Jaeckel:2010ni,Marsh:2015xka}. 

A spin-0 field with scalar and pseudoscalar couplings, as can arise for
axions in the presence of additional $CP$ violation, generates a parity-violating and
time reversal-violating monopole-dipole interaction between a polarized and
an unpolarized body~\cite{Moody:1984ba,Dobrescu:2006au,Fadeev:2018rfl,Okawa:2021fto,Cong:2024qly}.
This   is intriguing experimentally since only one body needs to be
polarized and the macroscopic force reverses sign under spin change 
$\mathbf S \rightarrow -\mathbf S$. Despite extensive atomic, molecular, torsion pendulum, and NV-center searches~\cite{Hoedl:2011zz,Rong:2017wzk,Cong:2024qly}, laboratory sensitivity to  monopole-dipole interactions remains limited at micrometer distances.

Precision Casimir experiments attain exceptional sensitivity to novel short range forces in exactly
the micrometer-scale regime~\cite{Mohideen:1998iz,Decca:2005qz,Chen:2014oda}, but have thus far constrained only spin-independent new interactions. 
Such isoelectronic and  
``Casimir-less'' measurements aiming to suppress the dominant Casimir vacuum background set leading limits
on short range Yukawa forces.
A variety of experimental Casimir geometry configurations allow complementary sensitive new force searches~\cite{Ma:2026afn}.
Since the test
bodies are unpolarized, the monopole-dipole contributions
vanish  in such setups after spin averaging. Constraints on pseudoscalar couplings from
unpolarized Casimir measurements have consequently relied on two-axion
exchange~\cite{Klimchitskaya:2020cnr}, which survives spin
averaging but is fourth order in the couplings~\cite{Adelberger:2003ph}. Polarizing one body enables
accessing second order monopole-dipole force.

At micrometer separations  spin-dependent force searches face several major challenges -  magnetic backgrounds from polarized sources and dominant Casimir and electrostatic forces. Torsion balance experiments employed magnetically compensated SmCo$_5$-AlNiCo spin sources~\cite{Heckel:2008hw,Heckel:2013ina}, but monopole-dipole searches were conducted at large millimeter gaps~\cite{Terrano:2015short}. However, at micrometer separations  local magnetization inhomogeneities can produce near field leakage, such as micron-scale domains in SmCo$_5$~\cite{Szmaja:2007smco}. When magnetic flux closes primarily within the plane of a spin-dominated source  the electron spins also lie predominantly parallel to the surface, which reduces monopole-dipole exotic signals. Proposed cantilever searches use oscillating patterned sources with magnetic flux return layers or superconducting shielding~\cite{Luo:2020monopoledipole,Wang:2026owg}, and garnet test mass proposals retain a conducting shield~\cite{Leslie:2014rig}. These layers increase the spin probe separation while ordinary forces correlated with the pattern can mimic the signal. Spin reversal with a levitated sensor suppresses spin-independent backgrounds, but residual magnetic forces limit its projected sensitivity~\cite{Xiong:2021spinmass}. Cantilever experiments have isolated velocity-dependent forces at micrometer gaps through distinct phases~\cite{Ding:2020mic,Li:2025ykh}, not applicable to monopole-dipole forces. Spin-spin searches requiring two polarized bodies have also been proposed using  cantilevers~\cite{Wang:2022bpw} and polarized Casimir plates~\cite{Bezerra:2016hjh}, although the projections using magnetized electron-spin plates are non-competitive.

Here we extend precision Casimir metrology from spin-independent to spin-dependent exotic force searches and put forth a pathway for sensitively probing short range monopole-dipole interactions. Unlike Casimir-less experiments that exploit material density contrasts~\cite{Decca:2005qz,Chen:2014oda}, our approach exploits spin contrasts. Reversing the polarization of one test body while preserving its mass distribution and electromagnetic surface response reverses the exotic force signal, while   Casimir, gravitational  and leading electrostatic forces remain as common  backgrounds. We introduce
 experiments based on materials with reversible spin polarization at vanishing net magnetization, which has been demonstrated for  the spin-orbital compensated ferromagnet Sm$_{1-x}$Gd$_x$Al$_2$~\cite{Adachi:1999zm,Adachi:2001compton,Taylor:2002sm,Qiao:2004xmcd}. Intrinsic compensation suppresses magnetic leakage without requiring bulky intervening shielding and thus  allows dense  normally polarized sources with closely separated test bodies.
We show that these setups could improve direct laboratory sensitivity to axion-mediated monopole-dipole forces by orders of magnitude over interaction ranges $\lambda\sim1-100~\mu{\rm m}$.

{\it Monopole-dipole forces and proximity scaling.-}
We consider static monopole-dipole interactions mediated by an axion particle with scalar and pseudoscalar  
couplings~\cite{Moody:1984ba,Fadeev:2018rfl,Pospelov:1998cpodd,
OHare:2020cpv,Okawa:2021fto,Dekens:2022gha}.
These interactions directly probe the product of a pseudoscalar
coupling and a \textit{CP}-violating scalar coupling,
complementing searches that constrain the individual couplings.
The pseudoscalar-scalar (ps) potential between two fermions $X$ and $Y$ is
\begin{equation}
V_{\rm ps}=-g_p^X g_s^Y \frac{1}{8\pi m_X}\hat{\mathbf{\sigma}}_X\cdot \hat{\mathbf r}\bigg(\frac{1}{r^2}+\frac{1}{\lambda r}\bigg)e^{-r/\lambda},
        \label{eq:V910}
\end{equation}
where $m_X$ is the mass of the polarized fermion $X$, $\hat{\mathbf{\sigma}}_X$ is its
Pauli spin operator, $r$ is the distance between the two fermions, $\hat{\mathbf r}$ is
the unit vector from $X$ to the unpolarized fermion $Y$, and $\lambda=m_\phi^{-1}$ is the
reduced Compton wavelength of the axion-type mediator of mass $m_\phi$. The couplings $g_p^X$ and
$g_s^Y$ are the pseudoscalar coupling of $X$ and the scalar coupling of $Y$. We use
natural units with $\hbar=c=1$, and quote observables in SI units.

We consider a standard Casimir experimental setup with two bodies. As a convention, we take both coated with Au, typical surface for precision Casimir metrology that does not oxidize, has well characterized dielectric
response  and equalizes the contact potentials
of the two bodies~\cite{Klimchitskaya:2009cw,Decca:2005qz,Chen:2014oda,Haghmoradi:2024ytw}.
Since $V_{\rm ps}\propto\hat{\mathbf\sigma}_X\cdot\nabla_Y(e^{-r/\lambda}/r)$, a
polarization field $\mathbf S(\mathbf x)$ (number density times spin direction) acts on
unpolarized matter like a Yukawa charge density $-\nabla\cdot\mathbf S$ plus a surface
density $\mathbf S\cdot\hat{\mathbf n}$, with $\hat{\mathbf n}$ the outward surface
normal. A slab with spin density $\rho_X$ polarized along its normal therefore carries
the charges $\pm\rho_X$ on its two faces.

\begin{figure}[t]
    \centering \includegraphics[width=0.99\linewidth]{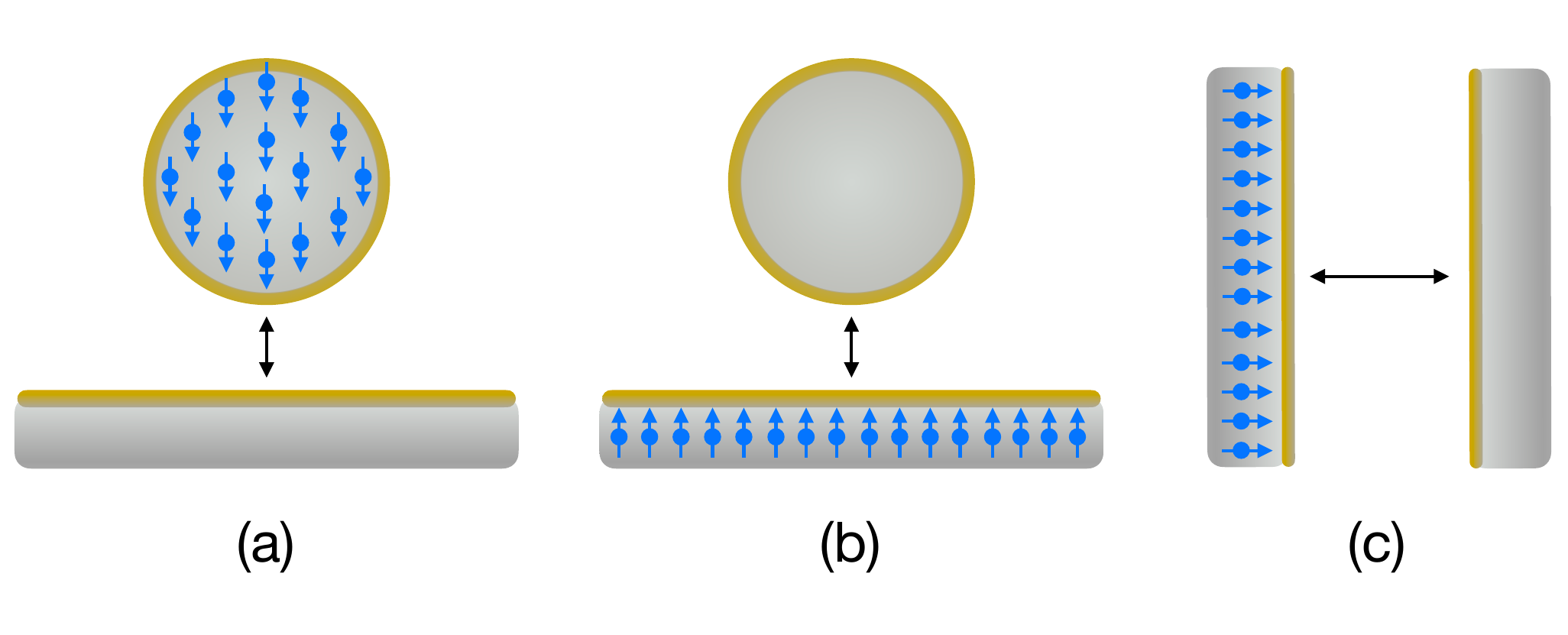}
    \caption{Considered precision Casimir   geometries with spin polarization. (a) Plate-sphere with a polarized source sphere. (b) Plate-sphere with a polarized source plate and an unpolarized sphere. (c) Parallel-plate with a polarized source plate. Blue arrows indicate electron spin polarization in magnetically compensated sources. Surfaces are taken to have an Au coating overlayer (gold).}
    \label{fig:schematic}
\end{figure}

In Fig.~\ref{fig:schematic} we illustrate  the considered experimental configurations. For two parallel plates of area $S$ with the unpolarized body containing fermions $Y$ at number density $\rho_Y$  the resulting monopole-dipole plate-plate force $F$ and pressure $P$ are  
\begin{equation}
    F_{\rm ps}^{\rm pl-pl} =   S P_{\rm ps}(\bar d_{\rm eff}) =
    \frac{\rho_X \rho_Y}{4 m_X}~g_p^X g_s^Y \lambda~
    e^{-\bar d_{\rm eff}/\lambda}~,
    \label{eq:PFA_plate_plate}
\end{equation} 
where  $\bar d_{\rm eff}=\bar d + t_{\rm Au}$ is the distance from the polarized source body material
to the passive body (the Au coating of the passive body is included as unpolarized scalar charge), $\bar d$ the  Au-Au surface separation, and $t_{\rm Au}$ is the
thickness of the considered Au coating layer on the polarized body that keeps the electromagnetic surface response  identical for distinct spin states.   
Finite homogeneous depths $a_X$ and $a_Y$ of the two bodies multiply
Eq.~\eqref{eq:PFA_plate_plate} by $(1-e^{-a_X/\lambda})(1-e^{-a_Y/\lambda})$.  

  For a plate-sphere system we employ proximity force approximation
(PFA), which treats the curved surface as a set of locally parallel plates (Supplemental
Material). For a sphere of radius $R$ the force is then $F=2\pi R\int_{\bar d_{\rm eff}}^\infty P_{\rm ps}(z)~dz$, giving
\begin{equation}
    F_{\rm ps}^{\rm pl-sph} = S_{\rm eff} P_{\rm ps}(\bar d_{\rm eff})
    = \frac{\pi R\lambda^2 \rho_X \rho_Y}{2 m_X}~g_p^X g_s^Y~ e^{-\bar d_{\rm eff}/\lambda}~,
    \label{eq:main-pfa-plate-sphere}
\end{equation}
where the sphere
acts as a plate with an effective area of $S_{\rm eff}=2\pi R\lambda$. 
Since the spin couples to the gradient of the Yukawa potential 
the plate-sphere monopole-dipole force scales as $\sim R\lambda^2$,
rather than typical $\sim R\lambda^3$ expected of spin-independent Yukawa interaction.
Further, since the signal scales with the spin source factor
$\rho_X(1-e^{-a_X/\lambda})e^{-\bar d_{\rm eff}/\lambda}$ it increases linearly with spin density but is
exponentially suppressed by additional separation or nonpolarized
layers modifying $d_{\rm eff}$. A source thinner than the interaction range 
$a_X\ll\lambda$ further suppresses the signal by $a_X/\lambda$.
Thus, dense and sufficiently thick spin sources without intervening
shields are favored. For a sphere-sphere system of radii $R_1$ and $R_2$, in Eq.~\eqref{eq:main-pfa-plate-sphere} $R$ is replaced by an effective radius $R_1R_2/(R_1+R_2)$.

For our analysis we compute the monopole-dipole force contributions fully numerically without assuming such approximations for finite geometries and including the contributions of Au coating. We verify the validity of PFA in Supplemental Material. The PFA applies for $\lambda,\bar d_{\rm eff}\ll R$ and lateral
dimensions much larger than $\lambda$.

{\it Differential spin measurements.-}
For configurations we consider that only the spin source is polarized while the passive body consists of a dense unpolarized material such as tungsten (W). Hence, permanent magnetic forces
and exotic dipole-dipole interactions between the two  bodies are absent.  For the source prepared in $(\pm)$ spin polarization
state, the total resulting force including monopole-dipole $F_{\rm ps}$  is
\begin{equation} 
F^{(\pm)}(\bar d)  = F_{\rm C}(\bar d) + F_{\rm G}(\bar d) + F_{\rm elec}^{(\pm)}(\bar d)   + F_{\rm mag}^{(\pm)}(\bar d) \pm F_{\rm ps}(\bar d_{\rm eff}),
\end{equation}
where $F_{\rm C}$, $F_{\rm G}$, $F_{\rm elec}^{(\pm)}$, and $F_{\rm mag}^{(\pm)}$
denote the Casimir, gravity, electrostatic (including patches), and residual magnetic
force contributions typically present in Casimir experiment measurements.

To sensitively probe exotic forces we combine  precision Casimir
metrology with differential spin measurements. We consider  magnetically compensated sources that can retain  a dense electron spin polarization at vanishing net
magnetization, such as Sm$_{1-x}$Gd$_x$Al$_2$~\cite{Adachi:1999zm,Adachi:2001compton,Taylor:2002sm,Qiao:2004xmcd}. Reversing the source's polarization changes the sign of the
monopole-dipole force, while the Casimir, gravitational  and
leading electrostatic forces are unchanged if the geometry and
electromagnetic surface response are preserved. Thus, common mode contributions cancel in
the difference between the reversed spin states.

Allowing for imperfect reproducibility and
correlated backgrounds after spin reversal, the force difference $D_F = F^{(+)}-F^{(-)}$ can be stated as
\begin{equation}
D_F  
\simeq 2F_{\rm ps}+F_C'\delta d
+\delta F_{\rm elec}+\delta F_{\rm mag}+\epsilon_F,
\end{equation}
where $\delta d=\bar d^{(+)}-\bar d^{(-)}$,
$F_C'=\partial F_C/\partial\bar d$,
$\delta F_i= F_i^{(+)}-F_i^{(-)}$, and $\epsilon_F = \sqrt{2}\Delta F_{\rm rand}$
is stochastic noise root mean square (RMS).
 We define the spin odd force estimate as
$\widehat F_{\rm odd}= D_F/2$ for differential spin measurements,
or the corresponding calibrated lock-in amplitude for periodic
modulation measurements, with uncertainty $\sigma_{F,\rm odd}$.
We consider for our projections that the separation variation between distinct spin
states is negligible with $\delta d \simeq 0$, as can be expected for example in case of in situ spin reversal. We further consider that spin-correlated backgrounds are below
the measurement threshold,  
motivated by density-modulated Yukawa force Casimir measurements~\cite{Chen:2014oda}.
For differential spin readout sensitivity, the systematic uncertainty requirement is
$|F_C'\delta d + \delta F_{\rm elec}+\delta F_{\rm mag}|/2
\lesssim\sigma_{F,\rm odd}$.

For our plate-sphere benchmark we consider $\bar d=0.3~\mu{\rm m}$, $R=149.3~\mu{\rm m}$ following Casimir experiment of Ref.~\cite{Chen:2014oda}. Rescaling the 
 measured Casimir force of Ref.~\cite{Chen:2014oda} gives
$F_{\rm C}\simeq11~{\rm pN}$ and
$|F_{\rm C}'|\simeq1.0\times10^{-4}~{\rm N~m^{-1}}$.
Since the Casimir force dominates signal at micron scales even small changes in separation for differential spin measurements can
produce a background. 
With $\sigma_{F,\rm odd}/F_{\rm C}\sim10^{-5}$,
we conservatively require $|F_{\rm C}'\delta d|
\lesssim\sigma_{F,\rm odd}$ that gives
$|\delta d|\lesssim1.2~{\rm pm}$. 
Such suppression   is supported
by the Casimir experiment  of
Ref.~\cite{Chen:2014oda}, which achieved  force contributions from surface unevenness below $\sim 0.05$~fN.

For plate-plate configuration benchmark at $\bar d=3~\mu{\rm m}$, following CANNEX experiment design~\cite{Haghmoradi:2024ytw}, the ideal metal
Casimir pressure is
$P_{\rm C}=\pi^2\hbar c/(240\bar d^4)
\simeq1.6\times10^{-5}~{\rm Pa}$.
The projected sensitivity
$\sigma_{P,\rm odd}/P_{\rm C}\sim6\times10^{-5}$
similarly gives the conservative requirement
$|\delta d|\lesssim50~{\rm pm}$.
The CANNEX design targets temperature stability of
$0.1~{\rm mK}$ and projection estimates separation variations of
$\sim5~{\rm pm}$ from thermal drift and
$\sim4~{\rm pm}$ from seismic motion~\cite{Haghmoradi:2024ytw},
indicating that the desired gap distance variation control is feasible
at fixed temperature provided $T_{\rm comp}$  is re-established after  preparation.
The static electrostatic patch pressure at $3~\mu{\rm m}$
is expected to be $10-40~{\rm nN~m^{-2}}$
~\cite{Sedmik:2021ngd}, requiring its spin-correlated
change to remain below $5-20\%$ for the desired
differential pressure sensitivity.

{\it Spin polarization source.-}
Sensitive monopole-dipole force searches require suppressing
magnetic backgrounds from polarized sources. Previous proposals
employing magnetic flux-return layers or superconducting
shielding~\cite{Luo:2020monopoledipole,Wang:2026owg} increase the separation between the polarized spins
and the probe. Instead, here we put forth a new strategy exploiting intrinsic
magnetic compensation to suppress magnetic backgrounds
at the source, without requiring an intervening shielding layer.
Notably, substantial electron spin polarization can
coexist with vanishing net magnetization when spin and
orbital magnetic moments can cancel
$\mathbf M=-\mu_B(g_{\rm spin}\langle\mathbf S\rangle+
g_{\rm orb}\langle\mathbf L\rangle)$,
where $\mu_B$ is the Bohr magneton,
$\langle\mathbf S\rangle$ and $\langle\mathbf L\rangle$
are the spin and orbital angular-momentum densities in
units of $\hbar$, and $g_{\rm spin}$ and $g_{\rm orb}$
are their respective  $g$-factors.
Cancellation between spin and orbital magnetic moments
allows $\mathbf M=0$ while retaining
$\langle\mathbf S\rangle\neq0$, as demonstrated in the
spin-orbital-compensated ferromagnet
Sm$_{1-x}$Gd$_x$Al$_2$ at its magnetization compensation
temperature~\cite{Adachi:1999zm,Adachi:2001compton,
Taylor:2002sm,Qiao:2004xmcd}.

\begin{figure*}[t]
\centering
\includegraphics[width=0.49\linewidth]{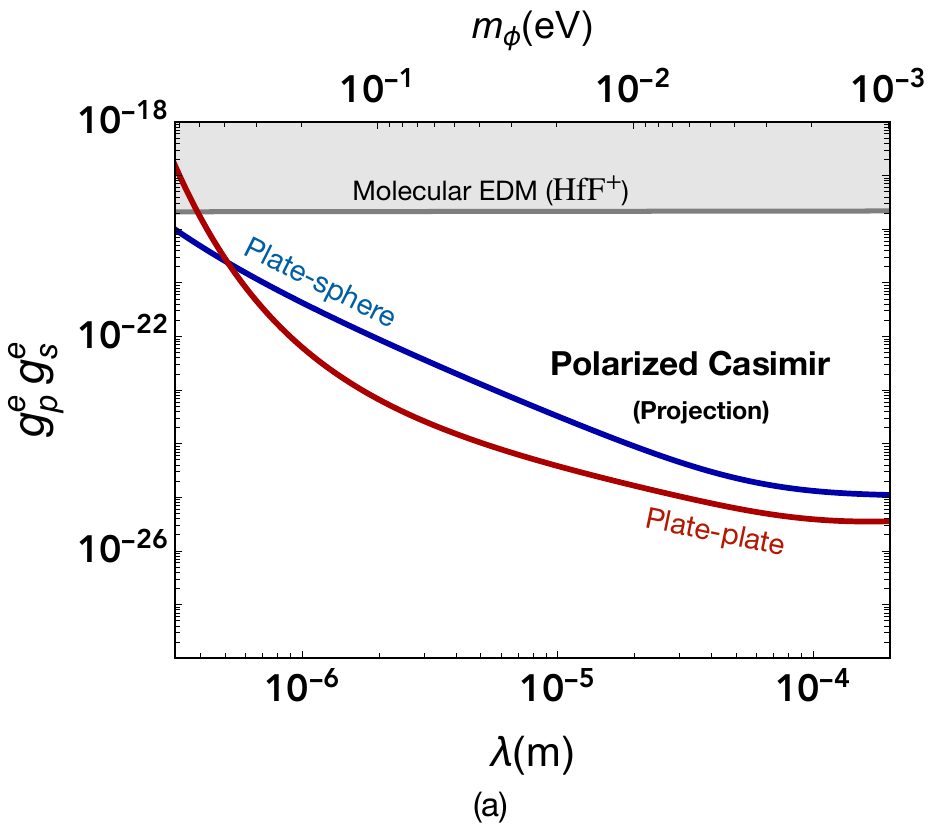}
\includegraphics[width=0.49\linewidth]{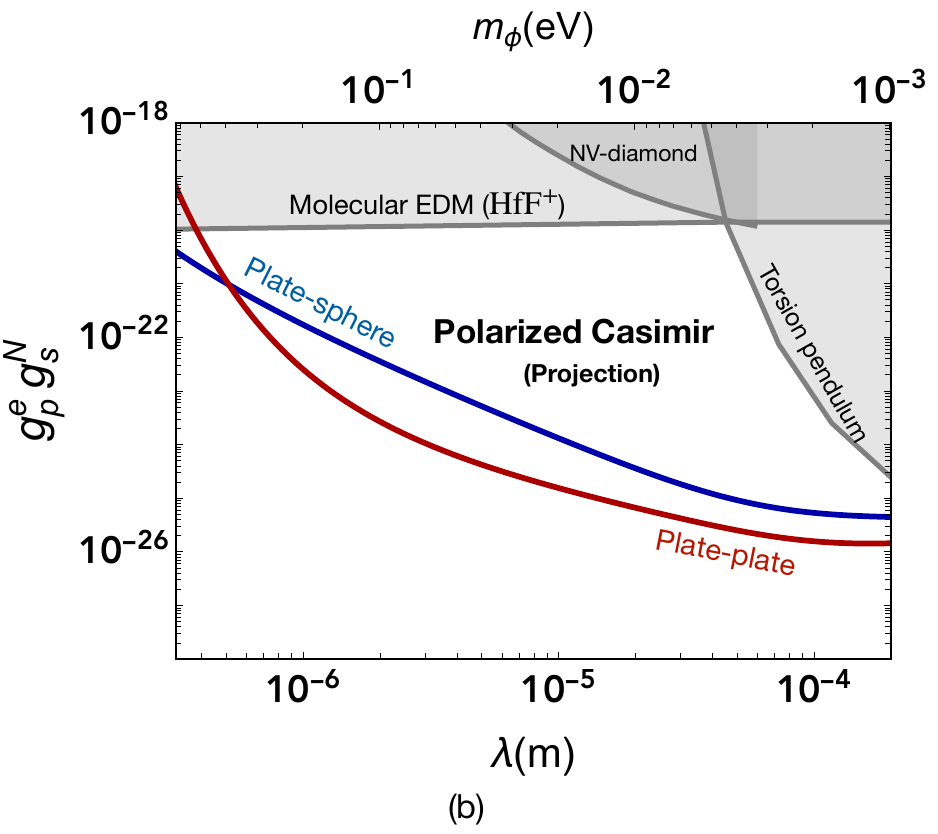}
\caption{Projected sensitivity to (a) $g_p^e g_s^e$ and (b) $g_p^e g_s^N$ new spin-force interactions. The plate-sphere curve is numerically calculated accounting for finite geometry for the periodically polarized rotating disk, using the experimental force threshold  of Ref.~\cite{Chen:2014oda}. The plate-plate curve assumes in situ spin reversal, using the CANNEX experiment design pressure target~\cite{Sedmik:2021ngd}. Spin-orbital compensated ferromagnet Sm$_{1-x}$Gd$_x$Al$_2$ source with $\rho_X=2.5\times10^{28}~{\rm m}^{-3}$ is considered, including Au coating. The constraints from laboratory molecular EDM~\cite{Prosnyak:2023duq}, NV-diamond~\cite{Liang:2022brr} and torsion-pendulum~\cite{Hoedl:2011zz} searches are shown.}
\label{fig:projected}
\end{figure*}

The proximity force scalings favor sources with high spin density $\rho_X$, minimal intervening unpolarized material layers  and suppressed local magnetization rather than merely a vanishing total magnetic moment. Although flux-closed SmCo$5$-AlNiCo assemblies used in torsion balance searches~\cite{Heckel:2008hw,Heckel:2013ina} provide high spin densities, local magnetic domain inhomogeneities can generate significant backgrounds at micrometer separations. For our plate-sphere benchmark, an assumed $\mathcal{O}(1\%)$ local magnetic-moment mismatch with correlation length $\xi=10~\mu{\rm m}$ results in an estimated magnetic force of $\sim1.3~{\rm fN}$ at the signal modulation harmonic for $\bar d_{\rm eff}=0.4~\mu{\rm m}$, about an order of magnitude above our force threshold (see Supplemental Material). In contrast, Sm$_{1-x}$Gd$_x$Al$_2$ exhibits compensation between spin and orbital magnetic moments at $T_{\rm comp}\simeq67-84~{\rm K}$, while retaining finite electron spin polarization confirmed by magnetic Compton scattering~\cite{Adachi:2001compton}. Residual magnetization arises from variations in composition and temperature, characterized by the RMS fluctuation $\sigma_M$ of the normal magnetization and its   correlation length $\xi$. We estimate a relative $\mathcal{O}(1\%)$ RMS variation in the Gd fraction $x$ gives $\mu_0\sigma_M\sim0.1-1~{\rm mT}$  for vacuum permeability $\mu_0$.

A stray magnetic field $B_{\rm leak}$ at the passive body of magnetic susceptibility  $\chi$  induces a force quadratic in $B_{\rm leak}$ that is even under
ideal reversal. Defining $\eta_{\rm rev}=|\delta F_{\rm mag}|/(2F_{\rm mag})$ as the fractional asymmetry that leaks into odd signal measurements, we estimate that a reference susceptibility $|\chi|\simeq 10^{-5}$  and near surface field
variation length of $0.3~\mu{\rm m}$ require
$\eta_{\rm rev}B_{\rm leak}^2\lesssim1.1\times10^{-7}~{\rm T}^2$ for the plate-sphere benchmark and
$\eta_{\rm rev}B_{\rm leak}^2\lesssim2.5\times10^{-10}~{\rm T}^2$ for the parallel-plate
benchmark. For a rotating Sm$_{1-x}$Gd$_x$Al$_2$ disk with a representative domain correlation length
$\xi=10~\mu{\rm m}$, the residual magnetization force at the signal harmonic is
$10^{-4}-10^{-2}~{\rm fN}$,   below the desired levels of around $\sigma_{F,\rm odd}=0.12~{\rm fN}$ adopted from Ref.~\cite{Chen:2014oda}, while the parallel-plate benchmark requires
$\eta_{\rm rev}\lesssim2.5\times10^{-4}$ at $B_{\rm leak}=1~{\rm mT}$. In Supplemental Material we discuss compensated
materials and  computations.  

Spin modulation in Casimir experiments can be implemented through several schemes. These include: (i) mechanical inversion (rotation) of a polarized sphere, requiring a sufficiently uniform surface to preserve ordinary forces; (ii) in situ spin reversal through temperature field preparation protocol, with both polarization states measured at the same compensation temperature~\cite{Adachi:2001compton}; (iii) periodic motion of polarized source, in analogy to torsion balance techniques~\cite{Terrano:2015short}. We adopt the rotating source approach (iii) for the plate-sphere benchmark and temperature field preparation (ii) for parallel plates.

For the plate-sphere setup we employ a rotating disk with periodically alternating spin polarization, replacing the material density contrast of conventional Casimir-less experiments  with spin contrast. The source consists of $n$ periodic regions each containing sectors of polarization magnitude $+P_0$ and $-P_0$, beneath
a continuous Au coating. Unlike the azimuthally polarized SmCo$_5$-AlNiCo
spin rings of torsion balance experiments~\cite{Terrano:2015short}, whose
monopole-dipole response arises from spin-density variations
at segment boundaries, our normally polarized sectors
retain the leading planar force response.

For a disk with periodic polarized regions rotating at angular frequency $\omega$ the signal
appears at $n\omega$, with first harmonic square wave coefficient
$\mathcal A_P=4/\pi$ and a PFA transfer factor describing signal suppression from neighboring oppositely polarized regions is
$\mathcal T_n=\exp[-n^2R\lambda/(2r_0^2)]$,
where $R$ is the sphere radius, $\lambda$ the interaction
range and $r_0$ the radial position of the measurement
track.
For $R=149.3~\mu{\rm m}$, $r_0=7.5~{\rm mm}$,
and characteristic $n=50$ we obtain $\mathcal T_{50}\simeq0.997$
at $\lambda=1~\mu{\rm m}$ and $0.967$ at
$10~\mu{\rm m}$. The parallel-plate benchmark instead uses full in place
spin reversal. For further details see Supplemental Material.

{\it Sensitivity to novel forces.-}
To determine the monopole-dipole force experimental reach we combine the above compensated source parameters and spin-modulation protocols together with detection thresholds. In the thick body PFA limit, 
requiring that the spin-odd amplitude exceeds given threshold value gives
\begin{align}
    g_p^X g_s^Y\Big|^{\rm pl-pl}
    &\gtrsim \sigma_{P,\rm odd}~\frac{4 m_X}{\rho_X \rho_Y \lambda}~e^{\bar d_{\rm eff}/\lambda},
    \label{eq:threshold-pp}\\
    g_p^X g_s^Y\Big|_{\rm pl-sph}
    &\gtrsim \sigma_{F,\rm odd}~\frac{m_X}{2\rho_X \rho_Y R\lambda^2\mathcal T_{n}(\lambda)}~e^{\bar d_{\rm eff}/\lambda},
    \label{eq:threshold-sp}
\end{align}
where Eq.~\eqref{eq:threshold-sp} includes $\mathcal A_P$ factor. For the plate-sphere projections we employ the experimental parameters of Ref.~\cite{Chen:2014oda}, namely $R=149.3~\mu{\rm m}$,
$\bar d=0.3~\mu{\rm m}$, and $\sigma_{F,\rm odd}=0.12~{\rm fN}$ for an integration time
$\tau\sim3000~{\rm s}$, and since that noise was set by spindle wobble rather than thermal noise we retain this threshold at   $T_{\rm comp}$. The
source is a $12.7~{\rm mm}$ radius Sm$_{1-x}$Gd$_x$Al$_2$ disk with
$\rho_X=2.5\times10^{28}~{\rm m}^{-3}$, $100~\mu{\rm m}$ active depth, and
$t_{\rm Au}=100~{\rm nm}$, so $\bar d_{\rm eff}=0.4~\mu{\rm m}$, and the passive body is an
Au-coated tungsten sphere. For the plate-plate projections we consider parameters of the CANNEX experiment design
$\sigma_{P,\rm odd}\sim1~{\rm nN~m^{-2}}$, integration time of $\tau  \sim 600~{\rm s}$ over
$\bar d=3-30~\mu{\rm m}$~\cite{Sedmik:2020cfj,Sedmik:2021ngd}, with the full
absolute pressure uncertainty as the sensitivity threshold. CANNEX is designed for room
temperature, but its sensor thermal noise $\sim40~{\rm pN~m^{-2}}$~\cite{Sedmik:2021ngd} 
is far below this target and we assume similar thresholds for a setup operated at
$T_{\rm comp}$. The plates have radius $5.872~{\rm mm}$, $200~\mu{\rm m}$ active depths,
and $100~{\rm nm}$ Au coatings. With a constant threshold our sensitivity envelope analysis selects
$\bar d=3~\mu{\rm m}$ at every $\lambda$. See Supplemental Material for details of computations and noise.

In Fig.~\ref{fig:projected} we display the projected sensitivity to monopole-dipole new forces. The plate-sphere and parallel-plate sensitivities can be compared using an effective pressure threshold for the sphere,
$\sigma_{F,\rm odd}/(2\pi R\lambda\mathcal A_P\mathcal T_n)$,
which puts Eq.~\eqref{eq:threshold-pp} and Eq.~\eqref{eq:threshold-sp} in similar form. For our benchmarks this corresponds to
$\simeq100~{\rm nN~m^{-2}}(1~\mu{\rm m}/\lambda)$,
around two orders of magnitude above the CANNEX pressure target at $\lambda=1~\mu{\rm m}$. However, the plate-sphere setup benefits from a smaller separation that enhances its signal relative to the parallel-plate setup by a factor of $e^{(3.1-0.4)~\mu{\rm m}/\lambda}$.

The high sensitivity of our projections arises primarily from the spin source geometry. At $\lambda=5~\mu{\rm m}$  our plate-sphere projections reach  $g_p^e g_s^N\simeq5\times10^{-24}$, improving by orders of magnitude upon the existing laboratory bound   of $\sim  10^{-20}$ from molecular EDM (HfF+) measurements~\cite{Prosnyak:2023duq} and the projected $9.4\times10^{-20}$ reach for a flux-closed magnetic source considering similar force thresholds~\cite{Wang:2026owg}. The significant improvement over the latter arises from the spin orientation, about two orders of magnitude, since in situ flux closure suppresses the planar response that our normally polarized source retains, and a comparable factor from the larger sphere, thicker source and absence of an intervening shield.

For $\lambda\gtrsim0.5~\mu$m our monopole-dipole projections for
direct Casimir searches significantly improve upon existing laboratory constraints. Indirect bounds inferred by
combining stellar cooling limits with stellar or laboratory limits on
the couplings can be considerably stronger~\cite{Fleury:2025ahw,Bottaro:2023gep}.
However, they rely on assumptions about stellar modeling as well as
environment-independent masses and couplings. In models with
environment-dependent masses or couplings, including chameleon-type
screening, particle production in stellar interiors can be strongly
suppressed thus relaxing these bounds by orders of magnitude relative to
laboratory searches~\cite{Jain:2005nh,Masso:2005ym,Khoury:2003aq,DeRocco:2020xdt,Terrano:2015short}.
In contrast, our approach enables direct controlled probes of these interactions  
independent of such assumptions.

{\it Conclusions.-}
Spin reversal offers a powerful route for searching for fundamental spin-dependent forces with precision Casimir experiments. Axion-mediated monopole-dipole forces change sign upon reversing the source polarization, while Casimir, gravitational  and electrostatic forces remain common mode to leading order enabling strong cancellations through differential measurements. We put forth magnetically compensated materials, exemplified by the spin-orbital-compensated ferromagnet Sm$_{1-x}$Gd$_x$Al$_2$, as spin sources combining  large electron spin densities with vanishing net magnetization. This intrinsically suppresses magnetic backgrounds without requiring intervening shielding, allowing spin-dependent force searches to exploit the small separations and high force sensitivity of precision Casimir metrology. Our plate-sphere and parallel-plate projections indicate possible improvements of up to five orders of magnitude over existing laboratory spin-force searches for $g_p^e g_s^{N}$ and $g_p^e g_s^{e}$ over regimes with $\lambda\sim1-100~\mu{\rm m}$. We outline the measurement protocols and requirements  needed to realize this experimentally. More broadly, our approach establishes spin polarization as a controllable tool in precision Casimir force metrology, opening new opportunities for direct laboratory tests of light bosons and fundamental spin-dependent interactions.

~\newline
\textbf{Acknowledgments.}
The work was supported by the World Premier International Research Center Initiative (WPI), MEXT, Japan. V.T. acknowledges support in part by the MEXT SPReAD Grant No. 26304205.

\setlength{\bibsep}{3pt}
\bibliography{references}

\clearpage
\newpage
\appendix
\onecolumngrid
\makeatletter
\def\@hangfrom@section#1#2#3{\@hangfrom{#1#2}#3}
\def\@sectioncntformat#1{\csname the#1\endcsname.\quad}
\def\thesubsection{\thesection.\arabic{subsection}}
\def\thesubsubsection{\thesubsection.\alph{subsubsection}}
\def\p@subsection{}
\def\p@subsubsection{}
\def\subsection{\@startsection{subsection}{2}{\z@}%
  {.8cm \@plus1ex \@minus .2ex}{.5cm}%
  {\normalfont\small\itshape\centering}}
\makeatother
\makeatletter
\@removefromreset{equation}{section}
\makeatother
\renewcommand{\theequation}{S.\arabic{equation}}
\setcounter{equation}{0}

\centerline{\large {Supplemental Material for}}
\medskip

{\centering\large\bfseries
Casimir Experiments as Spin-Force Detectors\par}
\medskip
{\centerline{Xiaolin Ma,  Volodymyr Takhistov, Hideo Iizuka}}
\bigskip
\bigskip

\vspace{1em}
\counterwithin{figure}{section}
\counterwithin{table}{section}

In this Supplemental Material we provide the force derivations, finite geometry calculations, modulation response and background estimates used in the main text. 

\FloatBarrier

\section{Point to Plate Monopole-Dipole Potential}
\label{sec:point-plate}

To compute spin-dependent macroscopic force effects between extended bodies we first calculate the monopole-dipole interaction between a polarized point fermion and an unpolarized plate. We then integrate this potential for two plates and use PFA for description of curved geometries  as in the case of plate-sphere system. 

\begin{figure}
    \centering
    \includegraphics[width=0.35\linewidth]{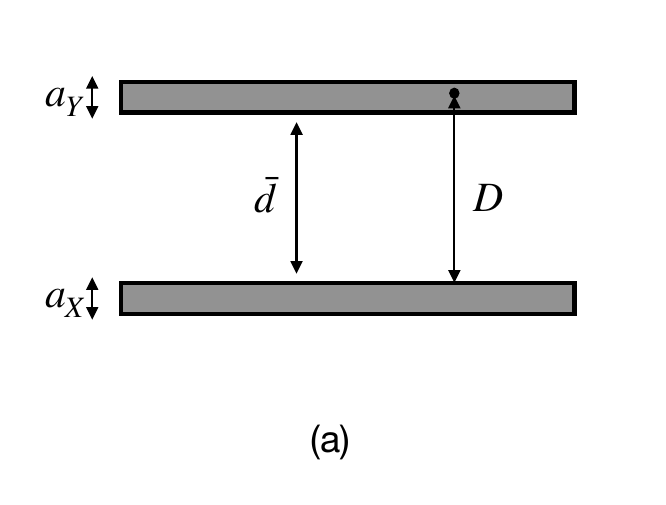}
    \hspace{4em}
    \includegraphics[width=0.35\linewidth]{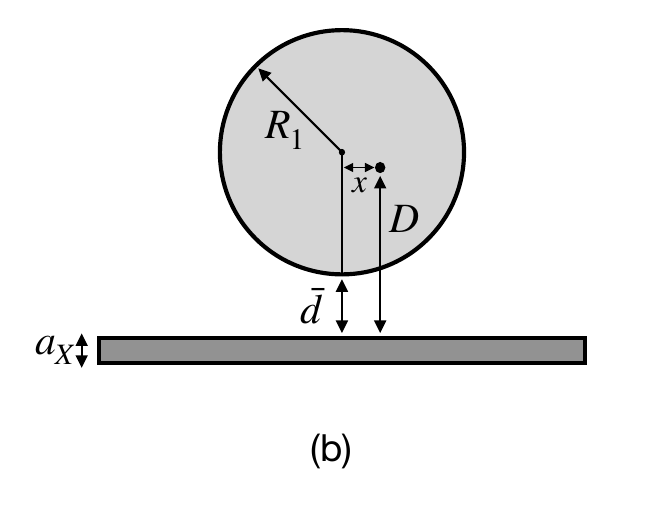} 
    \caption{Considered (a) plate-plate and (b) plate-sphere Casimir experiment geometries for spin force interactions.}
    \label{fig:casimir_1}
\end{figure}

We assign the pseudoscalar coupling to the polarized fermion $X$ with spin operator $\hat{\boldsymbol{\sigma}}_X$  and the scalar coupling to the unpolarized fermion $Y$. Exchange of a spin-0 boson with one pseudoscalar and one scalar vertex generates a parity-violating and time reversal-violating monopole-dipole potential~\cite{Moody:1984ba,Fadeev:2018rfl}. For a pseudoscalar axion the pseudoscalar coupling to matter fermions conserves CP, whereas a scalar coupling violates CP and can arise from additional CP-violating interactions~\cite{Pospelov:1998cpodd,OHare:2020cpv,Okawa:2021fto,Dekens:2022gha}. We parameterize the relevant spin-dependent terms as
\begin{align}
    V(m,n) \propto \rho_Y \int \cos^m(\theta)  \frac{e^{-d/\lambda}}{d^n}  \mathrm{d}V_Y,
\end{align}
where $d$ is the distance between the point particle and a volume element in the plate, and $\theta$ is the angle with respect to the surface normal. The integers $m$ and $n$ characterize the specific angular and radial dependence of the interaction. We consider that the spins are aligned perpendicular to the plate's surface, a configuration chosen to maximize the experimental sensitivity.

Consider a polarized point particle of body $X$ at the origin and an
unpolarized plate of body $Y$ with thickness $a_Y$ and number density
$\rho_Y$, whose near surface lies at $z=D$  so that the plate occupies
$D \le z \le D+a_Y$. In cylindrical coordinates $(r,\phi,z)$ the
distance to a volume element is $d=\sqrt{r^2+z^2}$ and the angle $\theta$
to the surface normal satisfies $\cos\theta = z/d$. After
integrating over $\phi$ the volume element is
$\mathrm{d}V_Y = 2\pi r \mathrm{d}r \mathrm{d}z$.

We obtain the total point-plate potential $V^{\rm p-pl}$ by integrating over the volume of the plate 
\begin{align}
    & V^{\rm p-pl}(D, a_Y, m, n) = \rho_Y \int_D^{D+a_Y} \mathrm{d}z \int_0^L 2\pi r   \bigg(\frac{z}{\sqrt{z^2+r^2}}\bigg)^m \frac{e^{-\sqrt{z^2+r^2}/\lambda}}{(\sqrt{z^2+r^2})^n} \mathrm{d}r,
\end{align}
where $L$ is the radius of the plate. We first evaluate the radial integral $I(z)$, considering definition of $d$, as
\begin{equation}
I(z) = \int_0^L 2\pi r \mathrm{d}r \left(\frac{z}{d}\right)^{m} \frac{e^{-d/\lambda}}{d^{ n}} = 2\pi z^{m} \int_{z}^{\sqrt{z^2+L^2}} d^{ 1-m-n} e^{-d/\lambda} \mathrm{d}d.
\end{equation} 

Assuming that  the plate radius is much larger than the interaction range and the separation with $L \gg \lambda, D$, the exponential factor $e^{-d/\lambda}$ allows us to extend the upper limit of integration to infinity with negligible error. Substituting $t = d/\lambda$, we obtain 
\begin{align}
    I(z) &\simeq 2\pi z^m \int_z^\infty d^{1-m-n} e^{-d/\lambda} \mathrm{d}d = 2\pi z^m  \lambda^{2-m-n} \int_{z/\lambda}^\infty t^{1-m-n} e^{-t} \mathrm{d}t \nonumber\\
    &= 2\pi z^m  \lambda^{2-m-n} \Gamma(2-m-n, z/\lambda).
\end{align}
where $L\to\infty$  and
$\Gamma(s,x)=\int_x^\infty t^{s-1}e^{-t} \mathrm{d}t$
is the upper incomplete gamma function. Using the generalized exponential integral $\mathrm{E}_n(x) =\int_1^\infty t^{-n}e^{-xt} \mathrm{d}t = x^{n-1} \Gamma(1-n, x)$, we can write the radial integral as
\begin{align}
    I(z) \simeq \begin{cases}
      \displaystyle \frac{2\pi}{z^{n-2}} \mathrm{E}_{m+n-1}(z/\lambda) & {\rm for }~ m+n\geq 2 \\
      \\
      2\pi z^{1-n} \lambda e^{-z/\lambda} & {\rm for }~ m+n=1.
    \end{cases}
    \label{eq:I_z}
\end{align}

To perform the remaining integration over $z$, we use the recurrence relations of  the
generalized exponential integrals  
\begin{align}
    \mathrm{E}_{n+1}'(x) = -\mathrm{E}_n(x), \quad \mathrm{E}_{n+1}(x) = \frac{1}{n}\left(e^{-x} - x \mathrm{E}_{n}(x)\right)~,
\end{align}
where  the prime
denotes the derivative with respect to $x$.
Integrating $I(z)$ with respect to $z$ yields 
\begin{align}
    \int    I(z) \mathrm{d}z=\begin{cases}
      \displaystyle -\frac{2\pi z^{3-n}\big[\mathrm{E}_{n-2}(z/\lambda)-\mathrm{E}_{m+n-1}(z/\lambda)\big]}{m+1},
       & \quad{\rm for }~ m+n\geq2,\\
      \\
      -2\pi z^{2-n}\lambda \mathrm{E}_{n-1}(z/\lambda),& \quad{\rm for }~ m+n=1.
   \end{cases}
\end{align}
Evaluating the definite integrals from $z=D$ to $z=D+a_Y$ gives the general expression for the point-plate potential 
\begin{align}
\begin{split}
    V^{\rm p-pl}(D, a_Y, m, n)
   ~&= \rho_Y \int_D^{D+a_Y} I(z)\mathrm{d}z \\
   &=
   \left\{
   \begin{aligned}
   &\dfrac{2\pi \rho_Y}{m+1}
   \Big[
   (D+a_Y)^{3-n}
   \Big(
   \mathrm{E}_{m+n-1}\!\left(\dfrac{D+a_Y}{\lambda}\right)
   - \mathrm{E}_{n-2}\!\left(\dfrac{D+a_Y}{\lambda}\right)
   \Big) \nonumber\\
   & 
   - D^{3-n}
   \Big(
   \mathrm{E}_{m+n-1}\!\left(\dfrac{D}{\lambda}\right)
   - \mathrm{E}_{n-2}\!\left(\dfrac{D}{\lambda}\right)
   \Big)
   \Big],
   && \quad{\rm for }~m+n\geq 2, \\[1.2ex]
   &2\pi\rho_Y \lambda
   \Big[
   D^{2-n}\mathrm{E}_{n-1}\!\left(\dfrac{D}{\lambda}\right)
   - (D+a_Y)^{2-n}\mathrm{E}_{n-1}\!\left(\dfrac{D+a_Y}{\lambda}\right)
   \Big],
   && \quad{\rm for }~m+n=1.
   \end{aligned}
   \right.\nonumber\\
   \nonumber\\
 \xrightarrow{\lambda,~D~\ll ~a_Y} &~~~\left\{
   \begin{aligned}
   &\dfrac{2\pi \rho_Y}{m+1}
   \Big[
   - D^{3-n}
   \Big(
   \mathrm{E}_{m+n-1}\!\left(\dfrac{D}{\lambda}\right)
   - \mathrm{E}_{n-2}\!\left(\dfrac{D}{\lambda}\right)
   \Big)
   \Big],
   && \quad{\rm for }~m+n\geq 2, \\[1.2ex]
   &2\pi\rho_Y \lambda
   \Big[
   D^{2-n}\mathrm{E}_{n-1}\!\left(\dfrac{D}{\lambda}\right)
   \Big],
   && \quad{\rm for }~m+n=1.
   \end{aligned}
   \right.
\end{split}
\end{align}

\section{Macroscopic Spin-dependent Force}
\label{sec:macro_force}
 
Starting from the point-plate monopole-dipole potential derived in Sec.~\ref{sec:point-plate}, we  can calculate the macroscopic forces for different geometries using PFA. Throughout $\bar d$ denotes the Au-Au surface separation, while $\bar d_{\rm eff}=\bar d+t_{\rm Au}$ includes the Au coating thickness $t_{\rm Au}$ on the polarized source. The passive body Au coating contributes to the scalar charge density and is included in the object density profile, rather than treated as an additional separation contribution. The PFA is valid in the short range regime $\lambda,\bar d_{\rm eff}\ll R$, where $R$ is the characteristic radius of the body surface curvature.
 
\subsection{Parallel-plate system} 
\label{sec:plate-plate}

We consider  parallel-plate system of common area $S$ that includes the polarized
source $X$ with thickness $a_X$ and spin density $\rho_X$, and the
unpolarized body $Y$ with thickness $a_Y$ and number density $\rho_Y$,
separated by $\bar d_{\rm eff}$ as depicted in \cref{fig:casimir_1}(a).
The total interaction potential follows from integrating the
point-plate result over plate $X$. A layer of $X$ at depth
$z\in[0,a_X]$ carries $\rho_X \mathrm{d}z$ polarized particles per
unit area and lies at distance $D=\bar d_{\rm eff}+z$ from the near
face of plate $Y$. Thus, one has
\begin{align}
    V_{\rm tot}(\bar d_{\rm eff}) = S\rho_X \int_0^{a_X}
    V^{\rm p-pl}(D, a_Y)\,\mathrm{d}z ,
\end{align}
where  $V^{\rm p-pl}(D,a_Y)$ is the
point-plate potential of Sec.~\ref{sec:point-plate}.

For the monopole-dipole potential in Eq.~\eqref{eq:V910} of the main text with the polarized spin along the surface normal and $\hat{\mathbf r}$ directed from $X$ to $Y$, the   terms combine to give
\begin{equation}
V_{\rm ps}^{\rm p-pl}(D,a_Y)
=-\frac{g_p^Xg_s^Y\rho_Y\lambda}{4m_X}
e^{-D/\lambda}\left(1-e^{-a_Y/\lambda}\right).
\label{eq:explicit-point-plate-sp}
\end{equation}
 The interaction energy per unit area
$W^{\rm pl-pl}({\bar d_{\rm eff}}) = V_{\rm tot}/S$ and the resulting force are
\begin{align}
    F^{\rm pl-pl}({\bar d_{\rm eff}}) =
    -\frac{\mathrm{d}V_{\rm tot}}{\mathrm{d}{\bar d_{\rm eff}}} =
    -S\frac{\mathrm{d}W^{\rm pl-pl}}{\mathrm{d}{\bar d_{\rm eff}}},
\end{align}
with the pressure $P({\bar d_{\rm eff}}) = -\mathrm{d}W^{\rm pl-pl}/\mathrm{d}{\bar d_{\rm eff}}$
giving the local interaction per unit area. The pressure $P(h)$ is the input to the PFA calculations
that follow.

Using the point-plate potential above, we obtain
the plate-plate interaction energy and pressure
magnitudes
\begin{align}\label{eq:finite-thickness-plate-plate}
\left|W^{\rm pl-pl}_{\rm ps}\right|
&=\frac{|g_p^Xg_s^Y|\rho_X\rho_Y\lambda^2}{4m_X}
e^{-\bar d_{\rm eff}/\lambda}\Theta_X(\lambda)\Theta_Y(\lambda),\\
\left|P^{\rm pl-pl}_{\rm ps}\right|
&=\frac{|g_p^Xg_s^Y|\rho_X\rho_Y\lambda}{4m_X}
e^{-\bar d_{\rm eff}/\lambda}\Theta_X(\lambda)\Theta_Y(\lambda), \notag
\end{align}
where $\Theta_i(\lambda)=1-e^{-a_i/\lambda}$.
The thick body limit corresponds to $a_X,a_Y\gg\lambda$, for which $\Theta_X\Theta_Y\to1$.

\subsection{Proximity force approximation for curved geometries}
\label{sec:pfa_results}

For curved geometries PFA
relates the interaction force to the parallel-plate
interaction energy per unit area, $W^{\rm pl-pl}(h)$.
Consider a sphere of radius $R$ above a plate with minimum
effective separation $\bar d_{\rm eff}$ between the polarized
source and passive body. At radial distance $x$ from the
point of closest approach  the local separation is
\begin{equation}
h(x)=\bar d_{\rm eff}+R-\sqrt{R^2-x^2}.
\end{equation}
Since the interaction decreases
exponentially with separation, for $x\ll R$ that dominate the     Yukawa-weighted $e^{-h/\lambda}$ integral when $\lambda,\bar d_{\rm eff}\ll R$ this reduces to parabolic approximation 
\begin{equation}
h(x)\simeq\bar d_{\rm eff}+\frac{x^2}{2R}.
\end{equation}
 
The PFA treats the sphere as a stack of concentric annuli. An annulus at radius $x$ has area $\mathrm{d}A = 2\pi x\mathrm{d}x$ and interacts with the plate at a local separation $h(x)$.
 The total force follows by integrating the local pressure, $P(h)=-\mathrm{d}W^{\rm pl-pl}/\mathrm{d}h$, over these annuli 
\begin{equation}
    F({\bar d_{\rm eff}}) = \int_{0}^{\infty} P\bigl(h(x)\bigr)\mathrm{d}A = \int_{0}^{\infty} 2\pi xP\Bigl({\bar d_{\rm eff}}+\frac{x^2}{2R}\Bigr)\mathrm{d}x.
\end{equation}
Changing the integration variable from $x$ to the local gap $h$, we have $\mathrm{d}h = (x/R)\mathrm{d}x$, which implies $\mathrm{d}A = 2\pi R\mathrm{d}h$. The integral becomes 
\begin{align}\label{eq:F_integral}
    F({\bar d_{\rm eff}}) &= 2\pi R \int_{{\bar d_{\rm eff}}}^{\infty} P(h)\mathrm{d}h = 2\pi R \int_{{\bar d_{\rm eff}}}^{\infty} \left(-\frac{\mathrm{d}W^{\rm pl-pl}}{\mathrm{d}h}\right)\mathrm{d}h.
\end{align}

Evaluating the integral yields 
\begin{align}
    F({\bar d_{\rm eff}}) &= -2\pi R \left[ W^{\rm pl-pl}(h) \right]_{{\bar d_{\rm eff}}}^{\infty} = 2\pi R \bigl[ W^{\rm pl-pl}({\bar d_{\rm eff}}) - W^{\rm pl-pl}(\infty) \bigr].
\end{align}
Assuming the interaction energy vanishes at infinite separation with $W^{\rm pl-pl}(\infty) \to 0$, we recover the standard PFA result 
\begin{equation}\label{eq:PFA_sphere_plate}
  F({\bar d_{\rm eff}}) = 2\pi RW^{\rm pl-pl}({\bar d_{\rm eff}}).
\end{equation}
Applying this relation to Eq.~\eqref{eq:finite-thickness-plate-plate} reproduces the plate-sphere result in Eq.~\eqref{eq:main-pfa-plate-sphere}.
Thus, the force between a sphere and a plate is $2\pi R$ times the interaction energy per unit area of parallel-plate system at separation ${\bar d_{\rm eff}}$. 

We extend the plate-plate result to a sphere of radius $R$
interacting with an infinite flat plate at a minimum gap ${\bar d_{\rm eff}}$ as shown in~\cref{fig:casimir_1}(b). Applying the PFA of
~Sec.~\ref{sec:pfa_results} in which the sphere surface is decomposed
into concentric annuli each treated as a local flat plate segment,
the total force is
\begin{equation}
    F^{\rm pl-sph}({\bar d_{\rm eff}}) = 2\pi RW^{\rm pl-pl}({\bar d_{\rm eff}}).
\end{equation}
The relation gives the plate-sphere force from the plate-plate energy density evaluated at the minimum
separation ${\bar d_{\rm eff}}$, with $R$ setting the geometric factor.

\subsection{Validity of proximity force approximation}
\label{ssec:pfa_validity}

The results above rest on the PFA that requires ${\bar d_{\rm eff}} \ll R$, and on the parabolic gap approximation that requires $x \ll R$ over the region contributing to the force. Since $P(h)$ decays exponentially in the excess gap $h-{\bar d_{\rm eff}} \simeq x^2/2R$ on the scale $\lambda$, the force integral is dominated by $x \lesssim \sqrt{2R\lambda}$, which satisfies $x \ll R$ whenever $\lambda \ll R$. The PFA is therefore highly accurate under the joint condition $\lambda, {\bar d_{\rm eff}} \ll R$, which defines the short range validity regime with subleading corrections suppressed by $O({\bar d_{\rm eff}}/R)$ or $O(\lambda/R)$. At larger $\lambda/R$, including the upper part of the displayed reach, we use the complete finite geometry numerical calculations.

\begin{figure*}[t]
\centering
\includegraphics[width=0.49\linewidth]{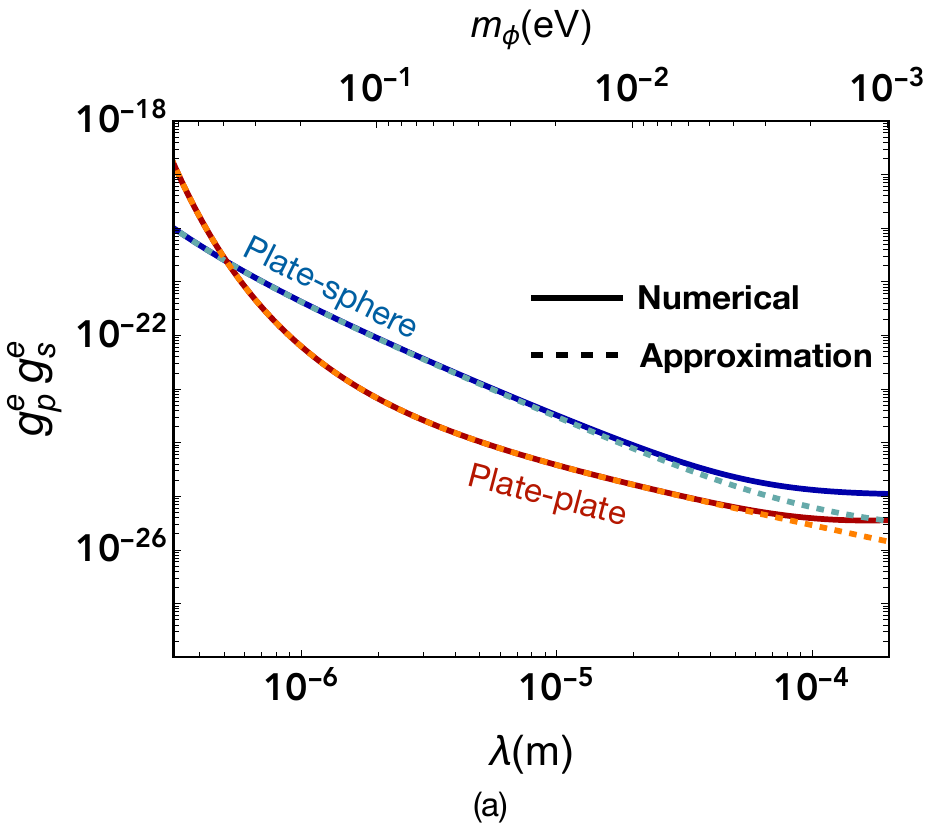}
\includegraphics[width=0.49\linewidth]{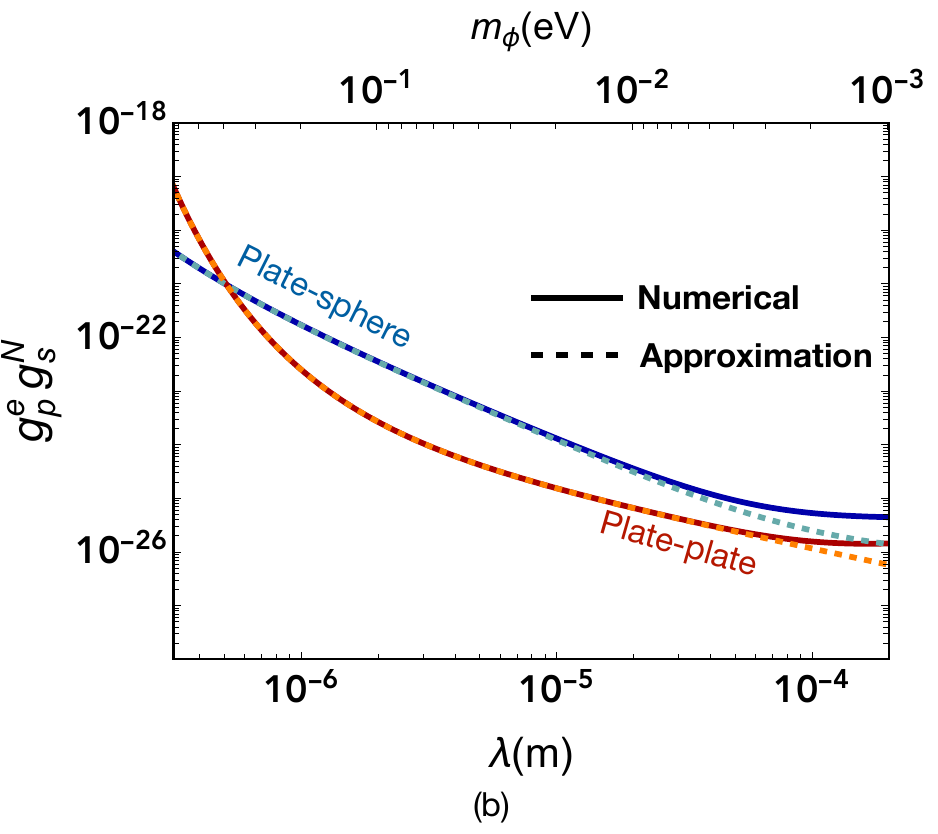}
\caption{Comparison of finite geometry numerical sensitivities (solid) with PFA estimates (dashed) for (a) $g_p^eg_s^e$ and (b) $g_p^eg_s^N$ spin force interactions in Casimir experiments. The blue curves denote the plate-sphere geometry. The red solid and orange dashed curves denote the plate-plate numerical and PFA results, respectively. The two calculations agree for $\lambda,\bar d_{\rm eff}\ll R_0$, where $R_0$ is the characteristic size of the source and test bodies. At larger $\lambda$, finite size and curvature effects lead to a deviation from the PFA scaling, as expected. 
}
\label{fig:placeholder}
\end{figure*}

For the comparison in Fig.~\ref{fig:placeholder}, the coating and finite depth factors are retained in the planar pressure. In natural units, its magnitude per coupling is
\begin{align}
\frac{|P_{\rm ps}^{(Y)}({\bar d_{\rm eff}})|}{|g_p^e g_s^Y|}
&=\frac{\rho_s\lambda}{4m_e}e^{-{\bar d_{\rm eff}}/\lambda}
(1-e^{-a_X/\lambda})  \left[\rho_Y^{\rm Au}(1-e^{-t'_{\rm Au}/\lambda})
+\rho_Y^{\rm W}e^{-{t'_{\rm Au}}/\lambda}(1-e^{-a_Y/\lambda})\right],
\label{eq:smx-pfa-layered-pressure}
\end{align}
where ${t_{\rm Au}}={t'_{\rm Au}}=100~{\rm nm}$ are the source and passive Au thicknesses,  respectively, and ${\bar d_{\rm eff}}=\bar d+t_{\rm Au}$. The plate-plate PFA uses infinite lateral planes and the thick body limit $a_X,a_Y\to\infty$, with {$\bar d=3~\mu{\rm m}$} and ${\bar d_{\rm eff}}=3.1~\mu{\rm m}$. Its numerical counterpart retains the $200~\mu{\rm m}$ active depths of Fig.~\ref{fig:projected}. For the plate-sphere PFA  we set $a_X=100~\mu{\rm m}$, take the passive half space limit $a_Y\to\infty$  and evaluate
\begin{equation}
F_{n,Y}^{\rm PFA}=\frac{4}{\pi}e^{-n^2R\lambda/(2r_0^2)}
~2\pi R\lambda~P_{\rm ps}^{(Y)}({\bar d_{\rm eff}}),
\label{eq:smx-pfa-sphere-force}
\end{equation}
at {$\bar d=0.3~\mu{\rm m}$} and ${\bar d_{\rm eff}}=0.4~\mu{\rm m}$. Thus, $\bar d_{\rm eff}$ accounts for the Au coating on the polarized source, while the passive side Au coating contributes to the scalar charge density without introducing an additional separation factor. In the homogeneous, thick body and unmodulated limits, these expressions reduce to Eq.~\eqref{eq:PFA_plate_plate} and Eq.~\eqref{eq:main-pfa-plate-sphere}.

\subsection{Direct numerical evaluation of monopole-dipole force}
\label{ssec:numerical_spin_force}

To test the validity of PFA and retain finite size effects, we integrate the
$z$-component of the pairwise force from Eq.~\eqref{eq:V910} over both
bodies 
\begin{equation}
    F_z (\bar d_{\rm eff})=
    -\int_{V_X}\mathrm{d}^3\mathbf{r}_1\int_{V_Y}\mathrm{d}^3\mathbf{r}_2~
    \rho_X(\mathbf{r}_1)\rho_Y(\mathbf{r}_2)~
    \pdv{}{z_{12}}V_{\rm ps}(\mathbf{r}_{12}),
    \label{eq:full-volume-z-force}
\end{equation}
where $\mathbf{r}_{12}=\mathbf{r}_2-\mathbf{r}_1$,
$z_{12}=z_2-z_1$. Here $\rho_X$ is the effective
spin density of the polarized body and $\rho_Y$ is the number density of
the unpolarized body. Spins are normal to the surfaces, and the overall
sign follows their orientation.

The derivative in Eq.~\eqref{eq:full-volume-z-force} is taken at fixed transverse separation and includes both the radial derivative and the derivative of $\hat{\boldsymbol\sigma}_X\!\cdot\!\hat{\mathbf r}$. For spins along $+\hat{\mathbf z}$, let $r=|\mathbf r_{12}|$ and $c=z_{12}/r$. The complete pairwise normal force is
\begin{align}
F_z^{\rm pair}
&=\frac{g_p^Xg_s^Y}{8\pi m_X}e^{-r/\lambda}\left[
(1-c^2)\left(\frac{1}{r^3}+\frac{1}{\lambda r^2}\right)
-c^2\left(\frac{2}{r^3}+\frac{2}{\lambda r^2}+\frac{1}{\lambda^2 r}\right)
\right]
\label{eq:complete-pairwise-z-force}
\end{align}
or  equivalently  $F_z^{\rm pair}=-(g_p^Xg_s^Y/8\pi m_X)~\partial_z^2(e^{-r/\lambda}/r)$. The full finite geometry calculation integrates this kernel over the sphere and compensated disk, accounting for the Au coatings, W substrate  and polarized source.

For Fig.~\ref{fig:placeholder}, the plate-sphere numerical curves use the rotating polarized plate and the unpolarized Au-coated W sphere of Fig.~\ref{fig:projected}. The comparison with the PFA uses the same $n=50$ modulation and $4/\pi$ first harmonic normalization. The numerical plate-plate calculation includes finite lateral dimensions and source depths, whereas the PFA assumes infinite lateral extent and thick bodies.
 We model the Au coatings, substrate  and polarized source as piecewise uniform regions, retaining their finite thicknesses and material-dependent densities. Translationally invariant directions are integrated analytically where possible, while the remaining plate-sphere integrals are evaluated numerically in coordinates centered on the point of closest approach.

The numerical and PFA results in Fig.~\ref{fig:placeholder} are seen to agree in the limit $\lambda,\bar d_{\rm eff}\ll R_0$, while finite size and curvature effects become significant at larger $\lambda$.

\subsection{Force response to periodic spin modulation}
\label{ssec:exact_lockin_response}

Lock-in readout is widely used in precision experiments to select the force component at a prescribed modulation frequency. We apply it here to spin modulation. Consider a fixed Au-coated passive sphere above a polarized
plate rotating at angular frequency $\omega$. We consider the polarized source to occupy $-a_X<z<0$, covered by an Au layer of thickness $t_{\rm Au}$. The sphere's closest point is at $z=\bar d_{\rm eff}=\bar d+t_{\rm Au}$, with its center at $z_c=\bar d_{\rm eff}+R$. We choose $x$ along the local direction of sector motion and use primed coordinates for source variables, denoting time by $\tau$. At the probe radius $r_0$, we define
\begin{equation}
 \psi_n(x',\tau)=k_nx'-n\omega\tau,
 \qquad
 k_n=\frac{n}{r_0}.
\label{eq:rotating-pattern-phase}
\end{equation}

We consider two square-wave polarization patterns, with alternating sectors of $+P_0$ and $-P_0$, or $P_0$ and $0$. Their spin densities are
\begin{equation}
 \rho_X^{\rm sq}(\mathbf r',\tau)
 =
 \rho_X
 \begin{cases}
 {\rm sgn}\!\left[\cos\psi_n(x',\tau)\right],
     & {\rm for }~+P_0/-P_0,\\[1mm]
 \left\{1+{\rm sgn}\!\left[\cos\psi_n(x',\tau)\right]\right\}/2,
     & {\rm for }~P_0/0,
 \end{cases}
\label{eq:periodic-square-wave-density}
\end{equation}
The component at the first lock-in harmonic is
\begin{equation}
 \rho_X^{(n)}(\mathbf r',\tau)
 =\mathcal A_P\rho_X\cos\psi_n(x',\tau),
 \qquad
 \mathcal A_P=
 \begin{cases}
 4/\pi,&{\rm for }~+P_0/-P_0,\\
 2/\pi,&{\rm for }~P_0/0.
 \end{cases}
 \label{eq:periodic-spin-density}
\end{equation}
For the modulating $P_0/-P_0$ source specified in the main text, $\mathcal A_P=4/\pi$. A $P_0/0$ pattern gives half this harmonic amplitude at the same $P_0$.
Here $\rho_X$ is the spin density at polarization $P_0$, and
$\mathcal A_P$ is the first Fourier coefficient of the square wave. Rotation
maps $k_n$ to the temporal frequency $n\omega$. 

At the reference phase
$\tau=0$, the PFA force follows from the local plate-plate pressure $P(h)$ 
\begin{equation}
 F_{n}^{\rm PFA}
 =
 \mathcal A_P
 \int \mathrm dx~\mathrm dy~
 P\!\left(h(x,y)\right)
 \cos(k_nx).
 \label{eq:periodic-density-pfa-integral}
\end{equation}
For a general spin-dependent interaction, the normalized PFA response factor due to modulation is
\begin{equation}
 \mathcal T_n^{\rm PFA}
 =
 \frac{\int \mathrm dx~\mathrm dy~
 P\!\left(\bar d_{\rm eff}+\frac{x^2+y^2}{2R}\right)\cos(k_nx)}
 {\int \mathrm dx~\mathrm dy~
 P\!\left(\bar d_{\rm eff}+\frac{x^2+y^2}{2R}\right)} .
\label{eq:periodic-density-pfa-general-response}
\end{equation}

Since the characteristic interaction region $\sim\sqrt{R\lambda}$ is much smaller than the probe radius $r_0$, the rotating polarization pattern can be treated locally as translating at speed $v=\omega r_0$. The first harmonic therefore oscillates at frequency $k_n v=n\omega$. By symmetry of the sphere's interaction region the sine component vanishes  so the force at $\tau=0$ gives the lock-in amplitude at $n\omega$, as considered above. The same argument applies to the exact volume integration below. Corrections from the local translation approximation are of order $\sqrt{R\lambda}/r_0$, while velocity-dependent effects are negligible.

Although Eq.~\eqref{eq:V910} contains explicit inverse powers of the 
separation between the sources, the transverse integral for an infinite planar source gives
$P(h)=K e^{-h/\lambda}$, with $K$ independent of $h$. Applying this local
pressure to the modulated plate-sphere geometry requires
$\lambda,\bar d_{\rm eff}\ll R$, so that the contributing patch has the parabolic gap
$h(x,y)\simeq\bar d_{\rm eff}+(x^2+y^2)/(2R)$ and can be treated as laterally infinite.
We also require $k_n\lambda\ll1$, so that the polarization pattern varies
slowly over the interaction range. Under these conditions,
Eq.~\eqref{eq:periodic-density-pfa-general-response} gives
\begin{equation}
 \mathcal T_n^{\rm PFA}
 =
 \frac{F_n^{\rm PFA}}{\mathcal A_PF_{k=0}^{\rm PFA}}
 =
 \frac{\int_{\mathbb R^2} \mathrm dx~\mathrm dy~
 e^{-(x^2+y^2)/(2R\lambda)}\cos(k_nx)}
 {\int_{\mathbb R^2} \mathrm dx~\mathrm dy~
 e^{-(x^2+y^2)/(2R\lambda)}}
 \simeq
 e^{-k_n^2R\lambda/2}.
 \label{eq:periodic-density-pfa-response}
\end{equation}
Accounting for the finite curvature, edge effects and
$k_n\lambda\gtrsim1$ requires complete volumetric integration. For the same periodic
density,
\begin{equation}
 F_{n}^{\rm exact}
 =
 -\int_{V_{\rm sph}}\mathrm d^3\mathbf r~\rho_Y
 \int_{V_{\rm pl}}\mathrm d^3\mathbf r'~
 \rho_X^{(n)}(\mathbf r')
 \pdv{}{z}V_{\rm ps}(\mathbf r-\mathbf r'),
 \label{eq:periodic-density-direct-integral}
\end{equation}
where $V_{\rm pl}$ denotes ${-a_X<z'<0}$. 

For an infinite periodic layer
and spins along $\hat{\mathbf z}$, Eq.~\eqref{eq:V910} becomes
\begin{equation}
 V_{\rm ps}(\mathbf r-\mathbf r')
 =
 \frac{g_p^Xg_s^Y}{8\pi m_X}
 \pdv{}{z}
 \left(\frac{e^{-|\mathbf r-\mathbf r'|/\lambda}}
 {|\mathbf r-\mathbf r'|}\right)
 .
 \label{eq:yukawa-gradient-form}
\end{equation}
Writing $\mathbf r=(\boldsymbol{\rho},z)$ and
$\mathbf r'=(\boldsymbol{\rho}',z')$, with
$\boldsymbol{\rho}=(x,y)$ and $\boldsymbol{\rho}'=(x',y')$ denoting
the lateral coordinates of the observation and source points, respectively,
the lateral transform is
\begin{equation}
 \int \mathrm d^2\boldsymbol{\rho}'~
 \cos(k_nx')
 \frac{e^{-|\mathbf r-\mathbf r'|/\lambda}}
 {|\mathbf r-\mathbf r'|}
 =
 \frac{2\pi}{q_n}e^{-q_n(z-z')}\cos(k_nx),
 \qquad
 q_n=\sqrt{\lambda^{-2}+k_n^2}.
 \label{eq:yukawa-lateral-transform}
\end{equation}

After integrating over the polarized depth, the force on a scalar
particle along the $z$ direction is
\begin{equation}
 F_{z,n}(x,z)
 =
 -\mathcal A_P
 \frac{g_p^Xg_s^Y\rho_X}{4m_X}
 {e^{-q_nz}}
 \left(1-e^{-q_na_X}\right)\cos(k_nx).
 \label{eq:periodic-point-force}
\end{equation}
The sign follows the spin orientation. For a homogeneous passive sphere of
scalar density $\rho_Y$, the remaining integral is
\begin{align}
 I_n
 &=
 \int_{\rm sphere}\mathrm d^3\mathbf r~
 e^{-q_nz}\cos(k_nx)=
 4\pi\lambda^3e^{-q_n(\bar d_{\rm eff}+R)}
 \left[
 \frac{R}{\lambda}\cosh\!\left(\frac{R}{\lambda}\right)-\sinh\!\left(\frac{R}{\lambda}\right)
 \right].
 \label{eq:finite-sphere-mode-integral}
\end{align}
Rotational invariance makes the centered sphere integral depend on the
complex wave vector $(ik_n,0,-q_n)$ only through
$q_n^2-k_n^2=\lambda^{-2}$. Combining the two integrals gives
\begin{align}
 \left|F_n^{\rm exact}\right|
 &=
 \mathcal A_P\left|g_p^Xg_s^Y\right|
 \frac{\pi\rho_X\rho_Y\lambda^3}{m_X}
 {e^{-q_n(\bar d_{\rm eff}+R)}}
 \left(1-e^{-q_na_X}\right)
 \left[
 \frac{R}{\lambda}\cosh\!\left(\frac{R}{\lambda}\right)-\sinh\!\left(\frac{R}{\lambda}\right)
 \right].
 \label{eq:exact-lockin-force}
\end{align}

Let $F_{k=0}^{\rm exact}$ denote the force from the corresponding uniformly
polarized plate. The normalized spatial response is
\begin{equation}
 \mathcal T_n^{\rm exact}(\lambda)
 =
 \frac{F_n^{\rm exact}}
 {\mathcal A_PF_{k=0}^{\rm exact}}
 =
 {e^{-(q_n-\lambda^{-1})(\bar d_{\rm eff}+R)}}
 \frac{1-e^{-q_na_X}}{1-e^{-a_X/\lambda}}
 .
 \label{eq:exact-sector-transfer}
\end{equation}
For $k_n\lambda\ll1$,
\begin{equation}
 q_n-\lambda^{-1}
 =\frac{k_n^2}{q_n+\lambda^{-1}}
 =\frac{k_n^2\lambda}{2}
 +\mathcal O(k_n^4\lambda^3).
\end{equation}
If the polarized layer is thick on the interaction scale and
{$\bar d_{\rm eff}\ll R$}, Eq.~\eqref{eq:exact-sector-transfer} reduces to
\begin{equation}
 \mathcal T_n^{\rm exact}
 \simeq
 \exp\left[-\frac{k_n^2\lambda}{2}
 {(\bar d_{\rm eff}+R)}\right]
 \simeq
 \exp\left[-\frac{k_n^2R\lambda}{2}\right].
 \label{eq:sector-transfer-pfa-limit}
\end{equation}

Thus, the PFA factor is recovered in the short range limit. In Eq.~\eqref{eq:exact-sector-transfer}, the exponential factor accounts for finite wavenumber propagation, while the second factor captures the finite polarized source depth. For a concentrically layered sphere  the normalized response remains unchanged, with the sphere form factor in Eq.~\eqref{eq:exact-lockin-force} replaced by a sum over shells. At larger $\lambda$, the complete $q_n$ dependence of Eq.~\eqref{eq:yukawa-lateral-transform} is required.

\section{Compensated Zero Magnetization and Spin Polarized  Materials}
\label{sec:spin_materials}

A central challenge in  spin-sensitive Casimir experiments is suppressing magnetic backgrounds while preserving sizable electron spin polarization relevant for the signal.  This is possible to achieve since the macroscopic magnetization $\mathbf{M}$ receives both spin and orbital contributions, allowing a nonzero electron spin polarization $\langle\mathbf{S}\rangle$ even when $\mathbf{M}=0$. Namely,
\begin{equation}
    \mathbf{M} = - \mu_B
    \bigl(g_{\rm spin}^{\rm L}\langle\mathbf{S}\rangle
    + g_{\rm orb}^{\rm L}\langle\mathbf{L}\rangle\bigr),
    \label{eq:magnetization-spin-orbital}
\end{equation}
where $g_{\rm spin}^{\rm L}$ and $g_{\rm orb}^{\rm L}$ are the spin and orbital g-factors, respectively. Here $\langle\mathbf S\rangle$ and $\langle\mathbf L\rangle$ denote the corresponding angular momentum densities divided by $\hbar$, so they have units of inverse volume and Eq.~\eqref{eq:magnetization-spin-orbital} has the dimensions of magnetization. 
Cancellation between spin and orbital contributions allows $\mathbf{M}=0$ while retaining $\langle\mathbf{S}\rangle\neq0$, suppressing magnetic backgrounds without eliminating sensitivity to exotic spin-dependent forces. We exploit this property in the proposed Casimir configurations.

Spin sources composed of magnetically compensated materials retain a net electron spin polarization while their net magnetic moment is suppressed. Their spin density sets the signal strength, while spatial variations of their residual magnetization set the near field magnetic background. One approach to achieve this is to combine  materials in which spin and orbital motion contribute differently to the magnetization, such as AlNiCo whose magnetization arises almost entirely from electron spin, and SmCo$_5$ where the Sm orbital moment contributes substantially~\cite{Heckel:2008hw,Heckel:2013ina}. Arranging them so that the magnetic flux closes within the assembly nearly cancels the net magnetic moment, while a net electron spin remains. However, the individual components remain magnetized, so the near field leakage depends on the source geometry and local magnetic structure~\cite{Heckel:2008hw}.
Compensation can also occur within a single material, as in chemically stable rare-earth iron garnets~\cite{Leslie:2014rig,Hughes:2025tbig}. In these ferrimagnets the rare-earth moment is opposed to the net iron moment, and the two cancel at the magnetization compensation temperature. Since the rare-earth moment includes an orbital contribution while the iron moment is almost entirely spin contribution, the cancellation of moments is incomplete, leaving a net electron spin polarization. For Dy$_3$Fe$_5$O$_{12}$ samples the measured net spin density at compensation is $\simeq 4\times10^{20} \hbar~{\rm cm}^{-3}$, corresponding to a polarized electron density $\rho_s\simeq8\times10^{20}~{\rm cm}^{-3}$~\cite{Leslie:2014rig}.

For our benchmark we consider the spin-orbital compensated ferromagnet Sm$_{1-x}$Gd$_x$Al$_2$ in which the Sm orbital moment cancels the combined Sm, Gd  and conduction electron spin moments at $T_{\rm comp}$~\cite{Adachi:1999zm,Qiao:2004xmcd}. Magnetic Compton scattering confirms a finite net spin polarization at compensation~\cite{Adachi:2001compton}. Since the total spin and orbital moments balance at compensation, the net polarized electron density is $\rho_s\simeq(1-x)(8/a^3)|\mu_L|/\mu_B$, where $8/a^3$ is the Sm site density of the SmAl$_2$ host with lattice constant $a\simeq0.794~{\rm nm}$~\cite{Avisou:2008zmf}. Taking the Sm orbital moment $|\mu_L|\sim1.5$-$2~\mu_B$/Sm measured by neutron diffraction below $60~{\rm K}$~\cite{Chatterji:2018sm}, and assuming it is retained in a fully polarized remanent state at $T_{\rm comp}$, results in $\rho_s\sim(2-3)\times10^{22}~{\rm cm}^{-3}$. The Curie temperature is $T_C\simeq127-128~{\rm K}$, while $T_{\rm comp}$ decreases from $84~{\rm K}$ at $x=0.018$ to $67~{\rm K}$ at $x=0.024$~\cite{Taylor:2002sm,Chatterji:2018sm}.

Near the compensation point  $M\simeq(dM/dT)(T-T_{\rm comp})$. For the background estimates we adopt $\mu_0|dM/dT|\sim0.2-2~{\rm mT~K^{-1}}$ as a representative range motivated by the measured magnetization curves~\cite{Adachi:2001compton,Taylor:2002sm}. The exact slope depends on the applied field and thermo-magnetic history of configuration. The two compositions quoted above give the finite difference estimate $|\Delta T_{\rm comp}/\Delta x|\simeq2.8\times10^3~{\rm K}$~\cite{Taylor:2002sm,Chatterji:2018sm}, which we use as an approximate $|dT_{\rm comp}/dx|$.  Temperature offsets and composition variations can generate residual local magnetization. At submicron gaps  magnetic domains and spatial variations of this residual magnetization could remain relevant even when the average moment vanishes. In Sec.~\ref{sec:microstructure} we give estimates for stray field and induced force, including the  uniformity requirements for Sm$_{1-x}$Gd$_x$Al$_2$.

Nuclear spin searches instead use hyper-polarized gas cells, such as $^3$He–$^{129}$Xe comagnetometers~\cite{Tullney:2013wqa,Bear:2000cd}. These gaseous sources have polarized spin densities of order $10^{17}~\mathrm{cm}^{-3}$, several orders of magnitude below those of solid electron spin sources and are incompatible with the rigid, planar geometry required for Casimir experiments. We therefore restrict our analysis to electron spin-dependent forces throughout this work.

In Tab.~\ref{tab:materials} we summarize  representative parameters of the unpolarized test bodies and candidate spin sources. The net polarized electron density sets the signal strength, while the residual magnetization variation, which we estimate in Sec.~\ref{sec:microstructure}, sets the near field magnetic background. Sm$_{1-x}$Gd$_x$Al$_2$ combines one of the highest spin densities with the smallest estimated residual magnetization, highlighting it as a favorable material for our setup.

\begin{table*}[t]
    \centering
    \small
    \setlength{\tabcolsep}{4pt}
    \begin{tabular}{|p{0.25\textwidth}|p{0.16\textwidth}|p{0.30\textwidth}|p{0.15\textwidth}|}
    \hline
        \textbf{Material} & \textbf{Polarization  fraction} & \textbf{Relevant density} $(\mrm{cm}^{-3})$ & \textbf{Estimated residual magnetization} $\mu_0\sigma_M$ (mT) \\
        \hline
        \multicolumn{4}{|l|}{\textbf{Unpolarized source/test bodies}} \\
        \hline
        Tungsten (W)~\cite{Kim_2020} & $P_N = 0$ & nucleon density $\rho_N \simeq 1.16\times10^{25}$, electron density $\rho_e \simeq 4.67\times10^{24}$~\cite{Kim_2020} & - \\
        \hline
        Gold (Au)~\cite{Kim_2020} & $P_N = 0$ & nucleon density $\rho_N \simeq 1.16\times10^{25}$, electron density $\rho_e \simeq 4.67\times10^{24}$~\cite{Kim_2020} & - \\
        \hline
        \multicolumn{4}{|l|}{\textbf{Polarized spin sources}} \\
        \hline
        {SmCo$_5$-AlNiCo~\cite{Heckel:2008hw}} & $P_e\sim10^{-2}$ (est.)~\cite{Heckel:2008hw} & {spin density contrast $|\rho_s^{\rm AlNiCo}-\rho_s^{\rm SmCo_5}|\simeq3.66\times10^{22}$~\cite{Heckel:2008hw}} & $10-100$ \\
        \hline
        Dy$_3$Fe$_5$O$_{12}$ at $T_{\rm comp}$~\cite{Leslie:2014rig} & $P_e\sim10^{-3}$ & electron spin density $\rho_s\simeq8\times10^{20}$~\cite{Leslie:2014rig} & $0.1-10$ \\
        \hline
        Tb$_3$Fe$_5$O$_{12}$ at $T_{\rm comp}$~\cite{Hughes:2025tbig} & $P_e\simeq2.6\times10^{-4}$ & electron spin density $\rho_s\simeq2.7\times10^{20}$ & $0.1-10$ \\
        \hline
        Dy$_6$Fe$_{23}$~\cite{Hou:2003zz} & $P_e\simeq1.2\times10^{-2}$ & electron spin density $\rho_s\simeq2.3\times10^{22}$ & $1-100$ \\
        \hline
        Gd$_{24}$Fe$_{66.5}$Co$_{9.5}$ (30 nm film, pre-pump)~\cite{Graves:2013nanoscale} & $P_e\sim1.4\times10^{-2}$ (est.) & Fe/Gd spin-only estimate $\rho_s\sim2.6\times10^{22}$ & $1-100$ \\
        \hline
        Sm$_{1-x}$Gd$_x$Al$_2$ at $T_{\rm comp}$~\cite{Qiao:2004xmcd,Taylor:2002sm} & {$P_e\sim0.02$ (est.)} & { $\rho_s\sim(2$-$3)\times10^{22}$~\cite{Chatterji:2018sm,Avisou:2008zmf}} & $0.1-1$ \\
        \hline
        Polarized ensemble ($^3\mathrm{He}$ + $^{129}\mathrm{Xe}$)~\cite{Tullney:2013wqa,Bear:2000cd} & $P_N\sim0.1-0.7$ & nuclear spin density $\rho_s \sim 10^{17}$~\cite{Tullney:2013wqa,Bear:2000cd} & - \\
    \hline
    \end{tabular}
    \caption{Representative parameters of the unpolarized test bodies and candidate spin sources for proposed Casimir-like spin force searches. $P_e$ ($P_N$) is the net polarized fraction of electrons (nuclei), and $\rho_s$ is the net polarized spin density. The residual magnetization $\mu_0\sigma_M$ is our estimate from the model of Sec.~\ref{sec:microstructure}, entries marked (est.) are likewise estimated.}
    \label{tab:materials}
\end{table*} 

\section{Spin Modulation Methods}

Several spin modulation approaches can isolate the monopole-dipole force while suppressing ordinary force backgrounds.

(i) Mechanical inversion: A polarized sphere with a uniform Au coating is rotated by $180^\circ$ about an axis transverse to the surface normal, reversing its spin projection relative to a passive plate. This produces the full polarization contrast $\Delta P_e=2P_e$, provided that the sphere's geometry and electromagnetic surface properties are sufficiently uniform to preserve the common mode forces.

(ii) In situ spin reversal: The spin polarization of a compensated source is reversed by preparing the material under an applied magnetic field at temperatures above and below its compensation temperature~\cite{Adachi:2001compton}. Both states are measured at the same temperature and separation, after removing the preparation field  without changing the geometry. Complete reversal gives $\Delta P_e=2P_e$.

(iii) Rotating spin patterned source: A fixed Au-coated sphere probes a rotating disk with alternating regions of opposite spin polarization  $+P_e$ and $-P_e$. Unlike the differential measurement approach for Casimir-less experiments in Ref.~\cite{Decca:2005qz,Chen:2014oda} we consider sources with spin contrast rather than material density contrast. A continuous Au coating overlayer and matched sector properties suppress spin-correlated Casimir, electrostatic  and gravitational backgrounds. For $n$ polarization periods, each consisting of two oppositely polarized sectors, rotation at angular frequency $\omega$ produces a signal at $n\omega$. A $P_e/0$ pattern yields half the first harmonic amplitude of the $+P_e/-P_e$ configuration.

More broadly, modulation between two  signed polarizations $P_{e,1}$ and $P_{e,2}$ gives a contrast $\Delta P_e=P_{e,1}-P_{e,2}$, compared with $\Delta P_e=2P_e$ for full polarization reversal. The resulting signal amplitudes, calibration requirements and spin correlated backgrounds are discussed below.

\subsection{Mechanical inversion}

Sensitivity of mechanical inversion (rotation) relies on uniformity of source geometry and electromagnetic surface properties. For a spherical source, perfect spherical symmetry is only approximate. Antipodal differences in radius and roughness change the gap and Casimir force, while differences in surface potential change the patch force. Radial runout, wobble induced gap distance changes and  odd patch force components should be measured at the modulation frequency, if the setup has periodic motion. The nominal inversion angle is $180^\circ$. An angular registration error $\delta\theta$ changes the normal spin projection by order $\delta\theta^2$ for an initially normal spin direction. Surface anisotropy can instead produce a background change linear in $\delta\theta$, which enters the measured force budget. The benchmark registration tolerance $\delta\theta\sim10^{-2}~{\rm rad}\simeq0.6^\circ$ is conservative relative to the few microradian angular errors resolved by precision rotary stage metrology~\cite{Xu_2014,Zhao_2021}. The relevant gap change must nevertheless be measured in the assembled apparatus, with an electrostatic gap calibration after each inversion.

For a transverse $180^\circ$ sphere inversion, a gap component $\delta d_n$ at the spin harmonic leaks the ordinary Casimir force into the signal channel as
\begin{equation}
|\delta F_{{\rm C},n}|\simeq
\left|\frac{\partial F_{\rm C}}{\partial d}\right||\delta d_n|
\leq \sigma_{F,\rm odd}.
\label{eq:casimir-gap-leakage}
\end{equation}
Ref.~\cite{Chen:2014oda} measured $F_{\rm C}=34~{\rm pN}$ at $d=200~{\rm nm}$ with a local power-law exponent $|\partial\ln F_{\rm C}/\partial\ln d|=2.78$. Rescaling the Casimir force to the $d=300~{\rm nm}$ separation used in the plate-sphere sensitivity calculation gives $F_{\rm C}\simeq11.0~{\rm pN}$ and $|\partial F_{\rm C}/\partial d|\simeq1.02\times10^{-4}~{\rm N~m^{-1}}$. Keeping the leakage below the adopted force threshold $\sigma_{F,\rm odd}=0.12~{\rm fN}$ therefore requires $|\delta d_n|<1.2~{\rm pm}$. Rotary stage runout and tilt measurements determine $\delta d_n$ directly~\cite{Xu_2014,Zhao_2021}.

\subsection{In situ reversal of compensated source}
\label{sec:inplace-reversal}

In situ modulation reverses the source spin without rotating or translating the source body. A concrete implementation uses thermally assisted state preparation in Sm$_{1-x}$Gd$_x$Al$_2$. Below $T_{\rm comp}$, the orbital contribution dominates the total magnetization, so the spin magnetization points opposite to an applied field that aligns the total moment. Above $T_{\rm comp}$, the spin contribution dominates and points along that field. Applying the same field direction on opposite sides of compensation prepares opposite spin orientations, which persist when the sample returns to $T_{\rm comp}$ at zero field~\cite{Adachi:2001compton,Avisou:2008zmf}.

The measurement cycle has two branches:
\begin{enumerate}
\item Set the source to a preparation temperature $T_-<T_{\rm comp}$ and apply a magnetic field along the surface normal to set its magnetic orientation. Return to $T_{\rm comp}$ while reducing the field to zero. After thermal equilibration and restoration of the measurement gap, record the force.
\item Set the source to $T_+>T_{\rm comp}$ and apply a field in the same direction to prepare the opposite spin orientation. Return to the same $T_{\rm comp}$, reduce the field to zero, restore the same gap, and record the force again.
\end{enumerate}
Both preparation temperatures are below the Curie temperature. Opposite spin states prepared through these two temperature-field histories have been observed by magnetic Compton scattering at a common $T_{\rm comp}$ and zero applied field~\cite{Adachi:2001compton}.

Repeating the two branches produces a sequence of opposite spin projections at fixed readout temperature, gap, and Au surface. For equal and opposite polarizations, the monopole-dipole force changes sign between the two readout states, giving the normalized difference $\widehat F_{\rm odd}=[F^{(+)}-F^{(-)}]/2=F_{\rm ps}$ in the ideal limit. The leading Casimir and gravitational forces are common mode. Force data are acquired during the equilibrated zero field intervals. The preparation and settling times set the modulation period.
 Spin correlated magnetic and electrostatic residuals enter the difference. 

\subsection{Periodic rotation of polarized patterned disk}

The rotating patterned source leverages the demonstrated Casimir-less
architecture of Refs.~\cite{Decca:2005qz,Chen:2014oda}, but instead of density contrasts incorporates spin polarization contrasts. Neighboring sectors
should have the same mass per area and share a continuous Au surface to
suppress ordinary force harmonics. A disk with $n$ polarization periods at
probe radius $r_0$, rotating at angular frequency $\omega$, produces the first
signal harmonic at $n\omega$. The square-wave patterns and their coefficients
$\mathcal A_P=4/\pi$ for $+P_0/-P_0$ and $2/\pi$ for $P_0/0$ are defined in
~\cref{eq:periodic-square-wave-density,eq:periodic-spin-density}. The PFA and
exact spatial responses are derived in
~\cref{eq:periodic-density-pfa-response,eq:exact-sector-transfer}. The measured
amplitude is
$F_n=\mathcal A_PF_{k=0}^{\rm exact}\mathcal T_n^{\rm exact}$.

Building on the experimental geometry of  Ref.~\cite{Chen:2014oda}, we take a sphere of radius
$R=149.3~\mu{\rm m}$ above a reference disk design with $n=50$ polarization
periods along a track of radius $r_0=7.5~{\rm mm}$. 
The finite wavenumber response $\mathcal T_{50}^{\rm exact}$ is evaluated
relative to the uniform $n=0$ source response, and the sensitivity projection
uses no other sector count.

 We adopt $\sigma_{F,\rm odd}= 0.12~{\rm fN}$  for $\tau \simeq 3000~{\rm s}$  following Ref.~\cite{Chen:2014oda}. Its measured topography contribution at the signal harmonic was below $0.05~{\rm fN}$ and its estimated patch contribution was below $10^{-17}~{\rm N}$~\cite{Chen:2014oda}. For a cryogenic implementation using Sm$_{1-x}$Gd$_x$Al$_2$ near $T_{\rm comp}$, we retain this force threshold as a benchmark, since the noise in Ref.~\cite{Chen:2014oda} was dominated by spindle wobble rather than oscillator thermal noise. A patterned spin disk at the same force target therefore requires
\begin{equation}
|\delta F_{{\rm topo},n}|<0.05~{\rm fN},\qquad
|\delta F_{{\rm patch},n}|<0.01~{\rm fN},\qquad
\eta_{\rm rev}F_{\rm mag}<0.1~{\rm fN},
\label{eq:modulation-background-targets}
\end{equation}
together with mass per area matching sufficient to keep the gravitational
harmonic below $0.1~{\rm fN}$. The first two bounds reproduce demonstrated
ordinary force controls, while the third constrains the compensated spin
source. These bounds constrain the individual backgrounds. The projection requires their combined contribution and the readout uncertainty to remain within the adopted $0.12~{\rm fN}$ budget. For the benchmark geometry, the $n=50$ Newtonian harmonic is reduced by $\simeq e^{-k_{50}R}\simeq0.4$ relative to the uniform $n=0$ source.

\section{Background Considerations}
\label{sec:experimental_implementation}

\subsection{Polarization calibration and backgrounds}

We choose the surface normal $\hat{\mathbf z}$ as a fixed quantization axis, and let each $P_i$ denote the signed polarization projected onto this axis. Thus $P_i>0$ and $P_i<0$ represent opposite orientations under the same direction convention. Let $F_0(\bar d_{\rm eff},\lambda)$ denote the force for a reference polarization $P_0>0$. For two polarization states $P_1$ and $P_2$, the spin-dependent force difference and polarization contrast are
\begin{equation}
\Delta F_{\rm sp}=\frac{P_1-P_2}{P_0}F_0,\qquad
\mathcal C_P=\frac{|P_1-P_2|}{2P_0},\qquad
|\Delta F_{\rm sp}|=2\mathcal C_P|F_0|.
\label{eq:polarization-contrast}
\end{equation}
Full reversal has $\mathcal C_P=1$, a $P_0/0$ sequence has $\mathcal C_P=1/2$, and two same sign polarization levels give the measured value $\mathcal C_P=|P_1-P_2|/(2P_0)$. Eq.~\eqref{eq:polarization-contrast} applies to an in situ flip, a plate-plate measurement  and a discrete sphere inversion. Each switching option should be tested and calibrated to measure $\mathcal C_P$, the residual magnetic moment, the surface potential  and the temperature in both polarization states.

For all spin modulation methods, the measured backgrounds are the components of ordinary forces that are correlated with the spin state. The absolute Casimir and Newtonian gravitational forces are common mode. For a sphere and a thick plate at $\bar d_{\rm eff}\ll R$, the static gravitational scale is $F_{\rm G}^{\rm sph-pl}\sim(8/3)\pi^2G\rho_{\rm sph}^{\rm mass}\rho_{\rm pl}^{\rm mass}R^3a_X$, yielding $F_{\rm G}^{\rm sph-pl}\simeq 9.38\times10^{-17}~{\rm N}$  for the active layer in our plate-sphere benchmark of $R=149.3~\mu\mathrm{m}$, active thickness $a_X=100~\mu\mathrm{m}$, tungsten sphere density $\rho_{\rm sph}^{\rm mass}\simeq1.93\times10^4~{\rm kg~m^{-3}}$, and an assumed composite plate density $\rho_{\rm pl}^{\rm mass}\simeq8.3\times10^3~{\rm kg~m^{-3}}$. The superscript distinguishes these mass densities from the spin number density $\rho_s$ used below.
For two parallel plates, the corresponding uniform Newtonian pressure is approximately $P_{\rm G}^{\rm pl-pl}=2\pi G\rho_1\rho_2 t_1t_2$ for bodies 1 and 2. Both uniform gravitational scales are independent of the spin orientation and therefore cancel in an ideal polarization-odd channel difference.

The patch requirement is $|\delta F_{\rm elec}|/2\lesssim\sigma_{F,\rm odd}$ for plate-sphere readout or $|\delta P_{\rm patch}|/2\lesssim\sigma_{P,\rm odd}$ for plate-plate readout. Mechanical sphere reversal tests antipodal differences of the patch map, while in-place spin reversal and a continuous-overlayer rotating disk keep the same Au surface facing the probe. Electrostatic calibration and Kelvin probe mapping can measure the remaining odd components~\cite{Speake:2003ik,Kim:2010zzp,Behunin:2012abc,Behunin:2012sphere,Garrett:2020qfr}.

For a compensated spin source acting on a passive scalar body of susceptibility $\chi$, the magnetic energy density $u_{\rm mag}$ and plate-sphere force scale as
\begin{align}
u_{\rm mag} \simeq \frac{|\chi|B_{\rm leak}^2}{2\mu_0},\quad \quad
F_{\rm mag}^{\rm sph-pl} \sim 2\pi Rl_Bu_{\rm mag},
\end{align}
where $l_B$ is the near surface variation scale. The leading induced force is even under the ideal transformation $\mathbf B_{\rm leak}\to-\mathbf B_{\rm leak}$ and thus is rejected in the odd channel difference. For two polarization state readout, we write the magnetic residual as $\eta_{\rm rev}=|\delta F_{\rm mag}|/(2F_{\rm mag})$, so that $|\delta F_{\rm mag}|/2=\eta_{\rm rev}F_{\rm mag}\lesssim\sigma_{F,\rm odd}$~\cite{Wang:2022bpw,Plumier:1983au}. 

For the sphere radius $R=149.3~\mu{\rm m}$, the local estimate evaluates to
\begin{equation}
F_{\rm mag}^{\rm sph-pl}\sim1.12\times10^{-25}~{\rm N}
\left(\frac{|\chi|}{10^{-5}}\right)\left(\frac{l_B}{0.3~\mu{\rm m}}\right)
\left(\frac{B_{\rm leak}}{10^{-8}~{\rm T}}\right)^2,
\label{eq:magnetic-benchmark}
\end{equation}
where the fiducial parameters represent a reference test body susceptibility
$|\chi| = 10^{-5}$ (for tungsten and gold the susceptibility is
$|\chi| \sim \text{few}\times 10^{-5}$, increasing the estimated force proportionally), a submicron surface field variation scale $l_B=0.3~\mu{\rm m}$, and the superconducting shielding benchmark $B_{\rm leak}=10^{-8}~{\rm T}$ simulated in Ref.~\cite{Wang:2022bpw}. In a compensated source disk, macroscopic moment compensation eliminates the long range dipolar field~\cite{Heckel:2008hw}, whereas near field leakage persists near sector boundaries and interfaces, governed by local magnetization mismatches and source geometry. Although the continuous Au overlayer fixes the dielectric boundary conditions for the Casimir force and screens electrostatic patch variations, it provides no attenuation of static magnetic fields. Magnetic suppression is decoupled from the surface force and relies strictly on intrinsic moment compensation together with external ambient field shielding. The spatial structure of this leakage and its dependence on the source microstructure are treated in Sec.~\ref{sec:microstructure}. The analogous plate-plate requirement is $\eta_{\rm rev}P_{\rm mag}\lesssim\sigma_{P,\rm odd}$, with $P_{\rm mag}\sim u_{\rm mag}$.

The force and pressure thresholds adopted in the main text require
\begin{align}
\eta_{\rm rev}B_{\rm leak}^2 &\lesssim
1.07\times10^{-7}~{\rm T}^2
\left(\frac{10^{-5}}{|\chi|}\right)\left(\frac{0.3~\mu{\rm m}}{l_B}\right)
&&\text{(sphere-plate)},\\
\eta_{\rm rev}B_{\rm leak}^2 &\lesssim
2.51\times10^{-10}~{\rm T}^2\left(\frac{10^{-5}}{|\chi|}\right)
&&\text{(plate-plate)}.
\label{eq:magnetic-inverse-budget}
\end{align}
For a leakage field $B_{\rm leak}=1~{\rm mT}$ and $|\chi|=10^{-5}$, these criteria require $\eta_{\rm rev}\lesssim0.11$  with $l_B=0.3~\mu{\rm m}$  for the plate-sphere setup and $\eta_{\rm rev}\lesssim2.5\times10^{-4}$ for the parallel-plate setup.  For a force-gradient readout, the gap derivative of $\delta F_{\rm mag}/2$ must remain below the calibrated force gradient uncertainty.

Even with $\eta_{\rm rev}=1$, the local estimate permits $B_{\rm leak}\lesssim0.33~{\rm mT}$ for plate-sphere readout and $B_{\rm leak}\lesssim16~\mu{\rm T}$ for plate-plate readout at the reference susceptibility and variation scale. The allowed field increases as $\eta_{\rm rev}^{-1/2}$ when reversal or harmonic rejection reduces the magnetic residual. The field and gradient at the probe position, measured in both spin states or over a full disk rotation, determine the spin correlated magnetic force.

Presence of a non-reversing ambient field $B_0$ gives a cross term $\sim |\chi|B_0B_{\rm leak}/\mu_0$ that follows the source reversal. Being linear in $B_0$, such contributions are readily suppressed by standard magnetic shielding and can be calibrated by applying a small known field.

\subsection{Local magnetic structure and near field leakage}
\label{sec:microstructure}

Since the monopole-dipole force is linear in spin, static spin inhomogeneity only rescales the signal through the interaction weighted mean spin density and does not generate a background. At submicron separations, the magnetic leakage is instead set by the residual magnetization and its spatial correlations, which arises for example from the presence of magnetic domains or composition variations. We express the field and induced force in terms of the normal magnetization variance $\sigma_M^2$ and correlation length $\xi$ and discuss their material dependent values.

Beyond the single field benchmark of Eq.~\eqref{eq:magnetic-benchmark}, lateral variations of the normal magnetization produce a magnetic patch distribution. One can consider $M_z(\mathbf x)=\bar M+\delta M(\mathbf x)$ with stationary isotropic covariance $\langle\delta M(\mathbf x)~\delta M(\mathbf x')\rangle=C(|\mathbf x-\mathbf x'|)$ and two-dimensional spectral density $S(k)=\int d^2r~C(r)~e^{-i\mathbf k\cdot\mathbf r}$. Away from edges, a laterally uniform ideal slab produces no exterior field, so the leakage is determined by $\delta M$. For a thick source whose magnetization variations are normal to the surface, the equivalent magnetic surface charge $\sigma_m=\delta M$ gives~\cite{Wallace:1951rec,Spector:1970aero}
\begin{equation}
\langle B_z^2\rangle(z)=\frac{\mu_0^2}{4}\int\frac{d^2k}{(2\pi)^2}~S(k)~e^{-2kz},\qquad \langle B^2\rangle=2\langle B_z^2\rangle .
\label{eq:microstructure-field}
\end{equation}
This is the magnetic analog of electrostatic patch potential models~\cite{Speake:2003ik,Behunin:2012abc}. For $C(r)=\sigma_M^2e^{-r/\xi}$, the spectrum is $S(k)=2\pi\sigma_M^2\xi^2/(1+k^2\xi^2)^{3/2}$. Thus $B_{z,\rm rms}\to\mu_0\sigma_M/2$ for $z\ll\xi$, while $B_{z,\rm rms}\simeq\mu_0\sigma_M\xi/(4z)$ for $z\gg\xi$. For example, for two state domains with normal magnetization $\pm M_s$ the variance is $\sigma_M^2=M_s^2(1-\bar m^2)$, where $\bar m=\bar M/M_s$ is the normalized mean magnetization~\cite{Hubert:1998dom}.

Integrating the induced energy density $-\chi B^2/(2\mu_0)$ over the sphere within the proximity force approximation yields the mean force magnitude
\begin{equation}
\bar F_{\rm mag}\simeq\frac{|\chi|}{2\mu_0}~2\pi R\int_{\bar d_{\rm eff}}^{\infty}\langle B^2\rangle(z)~dz=\frac{\pi}{4}~|\chi|~R~\frac{(\mu_0\sigma_M)^2}{\mu_0}~\xi~G\!\left(\frac{\bar d_{\rm eff}}{\xi}\right),
\qquad G(a)=\int_0^\infty\frac{e^{-2au}~du}{(1+u^2)^{3/2}},
\label{eq:microstructure-mean-force}
\end{equation}
where $\bar d_{\rm eff}=\bar d+t_{\rm Au}$ is the standoff from the magnetic surface, with the source Au coating adding to the gap and the passive body coating included in the magnetizable body. The PFA and thick source approximations assume $\xi\ll R$ and $\xi\ll a_X$, respectively. The function $G(a)$ approaches unity for $\xi\gg\bar d_{\rm eff}$ and $(2a)^{-1}$ for $\xi\ll\bar d_{\rm eff}$, giving $\bar F_{\rm mag}\propto\xi^2/\bar d_{\rm eff}$ in the latter limit. The gap filters spatial modes with $k\gtrsim\bar d_{\rm eff}^{-1}$, so domains or residual magnetization patches larger than the gap are not strongly suppressed.

Here we consider spin modulation by a rotating source disk as an example for requirements of the materials. Its fixed magnetic texture passes the probe and produces a background at the signal harmonic. Averaging over the interaction footprint and rotation harmonics gives, below the spatial roll-off,
\begin{equation}
F_{{\rm mag},n}\sim\bar F_{\rm mag}
\left(\frac{\xi}{2\pi R}\right)^{1/2}
\left(\frac{\xi}{2\pi r_0}\right)^{1/2},
\label{eq:microstructure-harmonic}
\end{equation}
where $r_0$ is the track radius. Requiring $F_{{\rm mag},n}\lesssim\sigma_{F,\rm odd}$ gives
\begin{equation}
\mu_0\sigma_{M,\max}\sim\left[
\frac{8\mu_0\sigma_{F,\rm odd}}
{|\chi|\xi^2G(\bar d_{\rm eff}/\xi)}
\sqrt{\frac{r_0}{R}}
\right]^{1/2}.
\label{eq:sigmaM-requirement}
\end{equation}

\begin{figure}[t]
\centering
\includegraphics[width=0.58\linewidth]{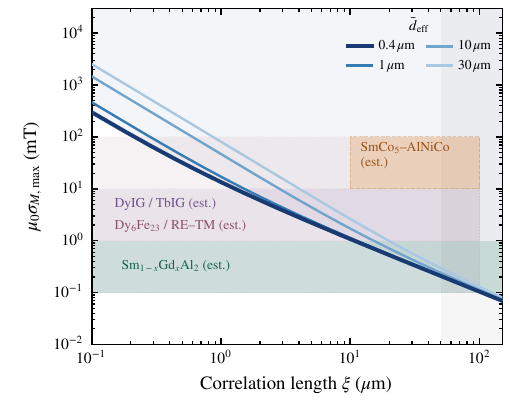}
\caption{Residual magnetization requirements for spin modulation with a rotating source disk for Casimir spin force experiments, from Eq.~\eqref{eq:sigmaM-requirement}. As characteristic parameters we adopt $|\chi|=10^{-5}$, $R=149.3~\mu{\rm m}$, $r_0=7.5~{\rm mm}$  and $\sigma_{F,\rm odd}=0.12~{\rm fN}$. The harmonic estimate includes spatial averaging without additional background suppression. The colored bands represent model estimates rather than measured bounds. The green band follows Eq.~\eqref{eq:smgdal-sigma}, using Sm$_{1-x}$Gd$_x$Al$_2$ magnetization data~\cite{Adachi:2001compton,Taylor:2002sm,Chatterji:2018sm}. The other bands use representative magnetization scales and assumed local mismatch fractions discussed in the text~\cite{Heckel:2008hw,Leslie:2014rig,Graves:2013nanoscale}. The faint shading above the most stringent curve indicates where the magnetic background exceeds the adopted force threshold. The gray region at $\xi\gtrsim50~\mu{\rm m}$ indicates the approximate limit of validity of the assumption $\xi\ll R,a_X$.}
\label{fig:sigmaM-requirement}
\end{figure}

In Fig.~\ref{fig:sigmaM-requirement} we demonstrate material bands estimated using $\mu_0\sigma_M=f_M\mu_0M_{\rm ref}$. For DyIG/TbIG~\cite{Leslie:2014rig,Hughes:2025tbig} and Dy$_6$Fe$_{23}$/rare-earth-transition metal (RE-TM) ferrimagnets~\cite{Hou:2003zz,Graves:2013nanoscale}, we adopt $\mu_0M_{\rm ref}=0.1-1~{\rm T}$, motivated by the DyIG sublattice moments $4.1$, $9.5$ and $13.6~\mu_B$ per formula unit at $T_{\rm comp}$, which give approximately $0.10$, $0.24$ and $0.34~{\rm T}$ at the sample density $3.4~{\rm g~cm^{-3}}$~\cite{Leslie:2014rig}. Taking representative local mismatch fractions $f_M=10^{-3}-10^{-2}$ for DyIG/TbIG and $10^{-2}-10^{-1}$ for Dy$_6$Fe$_{23}$/RE-TM gives $0.1-10$ and $1-100~{\rm mT}$, respectively, consistent with the $\sim1~{\rm mT}$-scale magnetization measured in TbIG within $10~{\rm K}$ of $T_{\rm comp}$~\cite{Hughes:2025tbig}. The correlation length is scanned over $\xi=0.1-100~\mu{\rm m}$. For SmCo$_5$-AlNiCo  a reference magnetization of order $1~{\rm T}$~\cite{Heckel:2008hw} and an assumed $1-10\%$ local mismatch give $10-100~{\rm mT}$. We scan $\xi=10-100~\mu{\rm m}$. 

Thermally demagnetized SmCo$_5$ exhibits domains of width $3-5~\mu{\rm m}$~\cite{Szmaja:2007smco}, larger than the benchmark gap. Taking $\xi=5~\mu{\rm m}$, $\mu_0M_s=1.07~{\rm T}$~\cite{Coey:2010mag}  and an illustrative remanence ratio $\bar m=0.9$, the binary domain model gives $\mu_0\sigma_M\simeq0.47~{\rm T}$ and $F_{{\rm mag},n}\sim6.7\times10^{-13}~{\rm N}$ at $\bar d_{\rm eff}=0.4~\mu{\rm m}$. SmCo$_5$-AlNiCo flux closure suppresses the far field~\cite{Heckel:2008hw,Terrano:2015short}, but a $1\%$ local magnetic-moment mismatch ($\mu_0\sigma_M\sim10~{\rm mT}$, $\xi=10~\mu{\rm m}$) gives $F_{{\rm mag},n}\sim1.3\times10^{-15}~{\rm N}$ at the same separation. Meeting the adopted force threshold requires the local magnetic moment mismatch to be below approximately $3\times10^{-3}$. Superconducting shielding reduces magnetic leakage~\cite{Wang:2026owg}, but an intervening shield of thickness $t_{\rm sc}$ suppresses the monopole-dipole signal by $e^{-t_{\rm sc}/\lambda}$.

Sm$_{1-x}$Gd$_x$Al$_2$ instead compensates spin and orbital magnetization locally near $T_{\rm comp}$, achieving an estimated $\rho_s\simeq(2$-$3)\times10^{22}~{\rm cm}^{-3}$, see Sec.~\ref{sec:spin_materials}. With $M\simeq(dM/dT)(T-T_{\rm comp})$ and composition fluctuations uncorrelated with domain orientation,
\begin{equation}
\sigma_M^2\simeq\left(\frac{dM}{dT}\right)^2\left[(1-\bar m^2)~\delta T_{\rm op}^2+\left(\frac{dT_{\rm comp}}{dx}\right)^2\sigma_x^2\right],
\label{eq:smgdal-sigma}
\end{equation}
where $\delta T_{\rm op}$ is the offset from the mean compensation temperature, $\sigma_x$ is the RMS Gd concentration variation  and $\bar m$ characterizes the domain orientation imbalance. Using the slopes in Sec.~\ref{sec:spin_materials}, $x\simeq0.02$ and $\sigma_x/x=1\%$ give $|dT_{\rm comp}/dx|\sigma_x\simeq0.56~{\rm K}$, corresponding to $\mu_0\sigma_M\simeq0.1-1~{\rm mT}$ at $\delta T_{\rm op}=0$. Once the mean temperature is tuned to compensation, temperature drifts of $1-10~{\rm mK}$ contribute at most $\mu_0\sigma_{M,T}\sim20~\mu{\rm T}$, subdominant to composition induced variations. For an illustrative correlation length $\xi=10~\mu{\rm m}$, this gives $F_{{\rm mag},n}\sim1.3\times10^{-19}$--$1.3\times10^{-17}~{\rm N}$, below the adopted force threshold. Since $\xi$ has not yet been experimentally determined, we extend the green band in Fig.~\ref{fig:sigmaM-requirement}   over the range of correlation lengths considered.

\section{Noise Budget for the Plate-Plate Geometry}
\label{sec:noise}

For plate-plate projections, we build our sensitivity on the detailed noise model of the
next generation CANNEX experiment~\cite{Klimchitskaya:2019fsm,Sedmik:2021ngd},
which targets a pressure sensitivity of
$\sigma_P\sim1~\mathrm{nN/m^2}$ between two parallel plates at
separations $d=3-30~\mu\mathrm{m}$. The measured pressure decomposes as
\begin{equation}
    \Ptot(d) = \Pcas(d) + \Ppatch(d) + P_{\rm mag}(d) + \Pnew(d) \pm \Delta P_{\rm tot}(d),
    \label{eq:measured-pressure}
\end{equation}
with $\Pcas$ the Casimir pressure, $\Ppatch$ the electrostatic
patch potential background, $P_{\rm mag}$ the residual magnetic pressure, $\Pnew$ the hypothetical new physics
signal, and $\Delta P_{\rm tot}$ the total uncertainty.

In the pressure channel, the signal is the static
extension $\Delta z_s$ of a micro-mechanical sensor read out by a
Fabry-P\'erot fiber interferometer,
$P = k\Delta z_s / S$~\cite{Sedmik:2021ngd}. Its sensitivity is
governed by the displacement resolution $\delta z_s$, which receives
contributions from the laser wavelength stability, detector voltage
noise, and residual vibrations.

At small separations, residual displacement noise $\delta d$ produces a nonlinear pressure bias,
\begin{equation}
    \Delta P_{\rm vib} \simeq
    \tfrac{1}{2}\langle\delta d^2\rangle|P''(d)|,
    \label{eq:vib-noise}
\end{equation}
which grows steeply for $d\lesssim5~\mu\mathrm{m}$.
The next generation
CANNEX design mitigates this with a two stage geometric anti-spring (GAS)
seismic isolation system derived from gravitational wave detector
technology~\cite{Sedmik:2021ngd}. The thermal Brownian noise of the
sensor, suppressed by a high quality factor $Q \sim 10^4$, contributes
only ${\sim}40~\mathrm{pN~m^{-2}}$ and is negligible relative to the
target sensitivity.

The leading deterministic electrostatic systematic is the residual
patch potential pressure. Microscopically, a metallic surface should
be treated as a spatially varying voltage map instead of  as a single
equipotential boundary. For two parallel plates this gives a pressure
set by the two-dimensional spectra of the surface-potential
correlations~\cite{Speake:2003ik,Kim:2010zzp,Behunin:2012abc},
\begin{align}
    |\Ppatch(d)| &\sim
    +\frac{\epsilon_0}{2}
    \int\frac{d^2{\bf k}}{(2\pi)^2}
    \frac{k^2}{\sinh^2(kd)}
    \left[C_{11}({\bf k})+C_{22}({\bf k})
    -2C_{12}({\bf k})\cosh(kd)\right].
    \label{eq:patch-spectrum-appendix}
\end{align}
Eq.~\eqref{eq:patch-spectrum-appendix} gives the pressure magnitude. The signed pressure is negative for an attractive interaction. Here
${\bf k}$ is the two-dimensional wave vector conjugate to the
coordinates along the plates, $k=|{\bf k}|$, so a patch mode with wave
number $k$ has lateral wavelength $2\pi/k$. The quantities $C_{11}$ and
$C_{22}$ are the voltage power spectra on the two plates, while
$C_{12}$ measures cross-correlations between the facing patch patterns.
The kernel shows that modes
with $kd\sim1$ dominate, short wavelength modes are exponentially
filtered by the gap, and long wavelength patches behave like a slowly
varying residual bias voltage. In the stochastic limit relevant for
CANNEX-like experiments, $\lambda_p\ll d$, where $\lambda_p$ is the characteristic lateral patch correlation length, and uncorrelated patches give the
short patch scaling law~\cite{Behunin:2012abc,Klimchitskaya:2019fsm,Sedmik:2021ngd}
\begin{equation}
    \Ppatch(d) \simeq
    \frac{3\zeta(3)\epsilon_0V_{\rm rms}^2
    \langle\ell^2\rangle}{4d^4},
    \label{eq:patch-pressure}
\end{equation}
with $V_{\rm rms}$ the RMS patch voltage and
$\langle\ell^2\rangle$ the second moment of the patch-size
distribution.

Approximating $\langle\ell^2\rangle\simeq\lambda_p^2$, the CANNEX
value $V_{\rm rms}\simeq 0.64~{\rm mV}$ gives
$P_{\rm patch}(8~\mu\mathrm{m})\sim0.2~\mathrm{nN/m^2}$ for
$\lambda_p\simeq0.5~\mu\mathrm{m}$. The same $d^{-4}$ scaling gives
$P_{\rm patch}(3~\mu\mathrm{m})\sim10$-$40~\mathrm{nN/m^2}$ for $\lambda_p\simeq0.5$-$1~\mu\mathrm{m}$, so patch backgrounds are mainly relevant at the
smallest accessible gaps and decrease rapidly at larger $d$. In the spin-modulated measurement proposed in
this work, the static part of $\Ppatch$ is common mode and cancels under an ideal polarization reversal. The quantity that enters the sensitivity
budget is therefore the odd residual
$\delta P_{\rm patch}=P_{\rm patch}^{(+)}-P_{\rm patch}^{(-)}$,
generated by state-dependent charging, contamination, or registration
changes between the two measurements. Equivalently, if such effects
convert only a fraction $\eta_{\rm patch}$ of the absolute patch
pressure into an odd component, one requires
$\eta_{\rm patch}|\Ppatch|/2\lesssim\sigma_{P,\rm odd}$ in the pressure channel. This residual can be measured
through electrostatic calibration and Kelvin probe mapping, and can be
reduced by in situ UV irradiation and Ar-ion cleaning, which suppress
large hydrocarbon- or water-induced patches~\cite{Sedmik:2021ngd}.

The same small patch result gives a useful benchmark for a
plate-sphere implementation. Within the PFA, valid
for $d\ll R$, the patch force is obtained by integrating the local
plate-plate pressure over the sphere curvature,
\begin{align}
    F_{\rm patch}^{\rm sph\text{-}pl}(d)
    &\simeq 2\pi R\int_d^\infty \Ppatch(z)~dz \simeq
    \frac{\pi\zeta(3)}{2}\epsilon_0 R V_{\rm rms}^2
    \frac{\lambda_p^2}{d^3},
    \qquad \lambda_p\ll d\ll R .
    \label{eq:patch-plate-sphere}
\end{align}
Considering the experimental parameters in Ref.~\cite{Chen:2014oda},
$R=149.3~\mu{\rm m}$ and $d=0.3-1~\mu\mathrm{m}$~\cite{Chen:2014oda},
and using the same illustrative $V_{\rm rms}=0.64~{\rm mV}$, this
becomes
\begin{equation}
    F_{\rm patch}^{\rm sph\text{-}pl}
    \simeq 0.38~{\rm fN}
    \left(\frac{R}{149.3~\mu{\rm m}}\right)
    \left(\frac{V_{\rm rms}}{0.64~{\rm mV}}\right)^2
    \left(\frac{\lambda_p}{0.1~\mu{\rm m}}\right)^2
    \left(\frac{0.3~\mu{\rm m}}{d}\right)^3 .
    \label{eq:patch-plate-sphere-number}
\end{equation}
Controlled short patches with $\lambda_p=0.05~\mu{\rm m}$
($0.1~\mu{\rm m}$) give
$F_{\rm patch}^{\rm sph\text{-}pl}\simeq0.095~{\rm fN}$
($0.38~{\rm fN}$) at $d=0.3~\mu{\rm m}$, falling to
$2.6\times10^{-3}~{\rm fN}$ ($1.0\times10^{-2}~{\rm fN}$) at
$d=1~\mu{\rm m}$. The corresponding reference scale is the adopted $\sigma_{F, {\rm odd}} = 0.12~{\rm fN}$~\cite{Chen:2014oda}. As in the plate-plate case, the ideal spin reversal rejects
the common mode patch force. The systematic entering the signal channel
is the odd residual
$\delta F_{\rm patch}=F_{\rm patch}^{(+)}-F_{\rm patch}^{(-)}$.

Surface roughness enters differently. The short distance plate-sphere
analyses indicate that roughness can be relevant at short
separations~\cite{PhysRevB.93.085434,van2007influence}. For the
present CANNEX range, $d=3$-$30~\mu{\rm m}$, short scale stochastic
roughness with $h_s\sim 1~{\rm nm}$ is negligible. The relevant
geometric systematic is instead the large scale waviness or sag of the
macroscopic plates, at the level of $\mathcal{O}(10~{\rm nm})$
peak to peak across the full force generating area. This shifts the
effective local gap entering $P(d)$, so the
appropriate treatment is to measure the plate topography, subtract the
known deformation profile, and include only the residual flatness
uncertainty in the small gap error budget~\cite{Sedmik:2021ngd}.

In a CANNEX-like experiment, the dominant noise source depends on the separation~\cite{Sedmik:2021ngd}. For
$d\gtrsim15~\mu\mathrm{m}$, the sensitivity is set by the interferometric
readout of the sensor extension. For
$d\lesssim5~\mu\mathrm{m}$, the main limitations are vibration nonlinearity,
residual patch effects, and the uncertainty in the large scale plate
flatness. The projected sensitivity is
$\sigma_P\sim1~\mathrm{nN~ m^{-2}}$~\cite{Sedmik:2020cfj,Sedmik:2021ngd}.
In the differential measurement proposed in the main text, the
spin-independent part of the Casimir and electrostatic backgrounds is
removed at the observable level. The relevant residual systematics are
therefore the odd components associated with patches, magnetic
backgrounds, and differential readout noise.

\end{document}